\documentclass[preprint,3p]{elsarticle}

\usepackage{amssymb}
\usepackage{amsmath}
\usepackage{bm}
\usepackage{soul}
\usepackage{subcaption}
\usepackage{xcolor}
\usepackage{pifont}
\usepackage{natbib}
\usepackage{graphicx}
\usepackage{dcolumn}

\journal{Nuclear Physics B}

\begin{document}

\begin{frontmatter}



\title{Sequential water wave reconstruction in VOF-based numerical wave tanks with the EnKF approach}

\author[1]{Liwen Yan}
\author[2]{Linyuan Che}
\author[1]{Jing Li}
\ead{lijing_@sjtu.edu.cn}  

\affiliation[1]{organization={Marine Numerical Experimental Center, 
State Key Laboratory of Ocean Engineering, 
School of Ocean and Civil Engineering, 
Shanghai Jiao Tong University}, 
addressline={}, 
city={Shanghai}, 
postcode={200240}, 
country={PR China}}

\affiliation[2]{organization={State Key Laboratory of Maritime Technology and Safety, 
Shanghai Ship and Shipping Research Institute Co., Ltd.}, 
city={Shanghai}, 
postcode={200131}, 
country={PR China}}

\begin{abstract}
Existing phase-resolved wave reconstruction methods are mostly based on potential flow theory, which limits their ability to capture strongly nonlinear phenomena such as wave breaking dynamics. 
In such cases, the importance of multiphase incompressible Navier--Stokes solvers becomes particularly evident.
These solvers form the foundation of the widely used numerical wave tanks in marine, ocean, and coastal engineering.
However, the high dimensionality of the state variables in such systems often hinders the efficiency of data assimilation.
The POD method is applied to reduce the order of the state vector in ensemble-based data assimilation, where instantaneous snapshots are naturally available.
Given the unsteady nature and random phase combinations of irregular waves, the effectiveness of single data assimilation is limited.
Sequential data assimilation is thus adopted to continuously update the wave profile as it evolves, ensuring that the wave field matches the real-world conditions.
Therefore, inflation is needed to mitigate the ensemble collapse.
Unlike single-phase flows without any interface, where the background error covariance can be inflated directly, to preserve physical constraints, when inflating for the phase field, the velocity field is simultaneously updated by incorporating potential flow formulas.
Three representative cases—--regular waves, irregular waves, and plunging waves—--are used to validate the proposed method and assess its performance.
The effect of assimilation parameters is also examined. 
The results demonstrate that the sequential data assimilation strategy achieves accurate free-surface reconstruction for a VOF-based numerical wave tank.
\end{abstract}

\begin{graphicalabstract}
\end{graphicalabstract}

\begin{highlights}
\item Developed a robust sequential data assimilation framework integrating the Ensemble Kalman Filter with a VOF-based Navier–Stokes solver for deterministic wave reconstruction.

\item Introduced a physics-constrained inflation strategy utilizing potential flow theory alongside POD-based order reduction to maintain physical consistency and computational efficiency.

\item Demonstrated the capability to capture complex, phase-resolved, non-linear phenomena—including irregular and plunging waves—that traditional potential flow-based methods fail to resolve.

\end{highlights}


\end{frontmatter}



\section{Introduction}
\label{sec:introduction}

Accurate simulation of water waves is underscored by the rapid growth of diverse industries, spanning from shipping and coastal engineering to offshore construction, aquaculture, and marine renewable energy.
Numerical wave tanks hence become an important platform for the study of wave dynamics and wave-structure interactions~\cite{peng2022numerical, hu2025hydrodynamic}.
While numerical wave tanks based on potential flow theory are valued for their computational efficiency, particularly in applications like design optimization~\cite{martin2020numerical}, high-fidelity viscous flow solvers remain indispensable in naval architecture and ocean engineering. 
This is due to their unique capability to simulate critical nonlinear phenomena, such as wave breaking, slamming, and green water~\cite{peng2022numerical}.
These high-fidelity simulations are typically grounded in computational fluid dynamics (CFD)~\cite{zhou2025development}, with the vast majority of solvers employing the Navier-Stokes equations or related formulations~\cite{peng2022numerical}. 
Algorithmically, these solvers are categorized into grid-based and particle-based methods~\cite{lyu2023derivation}. 
Within the dominant grid-based approaches, the finite volume method (FVM) is the most prevalent. 
However, as FVM requires explicit interface capturing for multiphase flows, specific techniques are employed, among which the volume of fluid (VOF) method is the most common in mainstream solvers~\cite{zhao2010numerical}, although the level-set method is also used~\cite{bihs2016new}.

The inherent discrepancies between computational models and physical reality necessitate rigorous validation against experimental benchmarks.
However, the validation of structural responses and loads under complex environments like irregular waves is typically confined to statistical agreements, which overlooks information on the individual deterministic occurrences.
This poses a significant challenge for reproducing critical but transient events, such as rogue waves, which can cause severe structural damage but are difficult to replicate deterministically in a numerical wave tank~\cite{nikolkina2011rogue}.
Even if the phases of wave components of the wavemaker are available, the actually generated wave spectrum often deviates from the target, given the limited wavemaker precision~\cite{katell2002accuracy}.
Despite the high degree of control in laboratory settings, the wave propagation is inevitably subject to various sources of error, such as boundary effects and tank reflections. 
These deviations are often difficult to measure directly or quantify accurately, resulting in a challenge to maintain consistency between experimental environments and numerical configurations. 
Furthermore, the numerical replication of phase-resolved waves that match those in field tests presents an even greater challenge, which is crucial for achieving high fidelity of numerical wave tank simulations and enhancing their practical engineering value.

Therefore, the phase-resolved wave reconstruction technique has garnered significant research interest.
Currently, methods used to solve this inverse problem mainly include data assimilation and machine learning.
Among the machine learning approaches, \citet{duan2020phase} proposed an artificial neural network-based wave prediction (ANN-WP) model for phase-resolved wave reconstruction and short-term forecasting. 
While their study included experimental validation, the model itself operates without embedding physical constraints.
The recent development of physics-informed neural networks (PINNs) \cite{jin2021nsfnets} provides a framework to incorporate prior knowledge of water wave physics directly into the training of neural networks. 
For instance, \citet{ehlers2025physics} employed PINNs to parameterize potential flow theory, establishing a computationally efficient method for wave prediction using only surface elevation measurements.
Machine learning, however, often relies on large amounts of training data, and the inherent lack of interpretability leads to weak generalization ability and fundamentally limits its practical application.

Data assimilation is a method that does not require pre-training and has strong physical consistency. 
Having been widely applied in fields such as meteorology and oceanography \cite{park2013data}, it also forms a major category in phase-resolved water wave reconstruction.
For example, \citet{zhu1994variational} combined a shallow water finite element model (FEM) with variational methods to invert the initial state of the model. 
Besides, \citet{wang2021phase} developed a coupled approach of the high-order spectral (HOS) method with the ensemble Kalman filter (EnKF), through which the elevation measurements can be incorporated into the simulation to improve the forecast performance.
To mitigate the impact of wave field disturbances in strongly nonlinear waves, \citet{wang2025nonlinear} proposed another model based on the pseudospectral Fourier-Legendre method with the EnKF, achieving enhanced predictive accuracy for surface elevation in cases of high wave steepness.
\citet{wang2024accelerating} conducted a numerical test on tsunami evolution by projecting the Saint-Venant equations onto a reduced-order space and combining them with the EnKF, revealing superior stability compared to the full-order model with EnKF. 
This example achieves wave reconstruction through the retrieval of initial conditions without changing boundary conditions.
These applications are still primarily based on potential flow theory or other simple wave models, having significant limitations when it comes to accurately capturing complex hydrodynamic behaviors. 

Although data assimilation has been successfully integrated with Navier-Stokes solvers for various flow problems, its application to water wave reconstruction in numerical wave tanks remains limited.
\citet{he2024four} recently applied four-dimensional variational data assimilation to reconstruct the super-temporal-resolution turbulent jet, where synthetic measurements are used from a large-eddy simulation.
\citet{zhang2022ensemble} incorporated Reynolds-averaged Navier--Stokes solver, EnKF, and neural networks altogether to learn a nonlinear eddy viscosity model.
This strategy shows that one turbulence model learned from one flow can predict flows in similar configurations.
Moreover, \citet{mons2016reconstruction} explored the unsteady inflow reconstruction in a flow around a cylinder by using different assimilation schemes.
Their group also discussed the optimal sensor placement for this unsteady flow problem~\cite{mons2017optimal}.
Besides, there are some studies dealing with the free surface as a boundary condition for Navier--Stokes solvers when using FEM.
For example, \citet{gejadze2006open} employed the free surface barotropic Navier-Stokes equations as the flow model, using velocity and elevation as observation data to determine boundary conditions.
\citet{fang2005adaptive} incorporated an adaptive mesh method into the adjoint model to reconstruct the free surface boundary of two-dimensional waves using elevation as the observation. 
However, it should be noted that FEM requires complex mesh deformation and remeshing when the interface undergoes large deformations, so it is hard to utilize FEM to deal with wave-structure interactions.
FVM thus is the mainstream for Navier-Stokes-based numerical wave tanks.
In contrast to potential flow theory, where state variables are limited to wave surface elevation and velocity potential, the Navier-Stokes solver governs the entire flow field as the state.
Furthermore, the use of multiphase schemes such as VOF introduces the fields with sharp gradients or even discontinuities near the interface, posing significant challenges for wave reconstruction.

The motivation of the present article is to demonstrate the feasibility of data assimilation for phase-resolved wave reconstruction in VOF-based numerical wave tanks, an area that remains largely unexplored.
To this end, we develop an integrated strategy to sequentially update the wave profile and simultaneously ensure the inflation is physically constrained.
Crucially, the inlet boundary conditions for irregular waves are non-periodic and stochastic, generating a unique wave realization that varies continuously in time.
This distinct feature imposes significant difficulty in simultaneously reconstructing the free surface and the underlying flow field, as the numerical model must constantly adapt to the unpredictable incoming waves rather than converging to a repetitive pattern.
This is why the sequential assimilation approach is implemented, enabling progressive correction of the wave profile as it propagates.
Within this sequential framework, inflation is introduced to counteract ensemble collapse.
In contrast to single-phase flows without interface that permit direct inflation of background error covariance, inflation of the phase field necessitates a corresponding update to the velocity field to satisfy physical constraints.
Based on the extracted Fourier components of the wave profile, the velocity field is updated by imposing the corresponding solution from linear water wave theory.
The EnKF~\cite{evensen1994sequential} is thus employed for the data assimilation, as it is both sequential and non-intrusive in nature.
POD is used to largely mitigate the computational cost associated with the gain matrix by reducing the system dimensionality. 
Besides, the structural information of flow fields may benefit the stability of the data assimilation process~\cite{li2025role}. 
Moreover, the ensemble-based assimilation framework naturally provides instantaneous snapshots, facilitating the construction of the POD basis. 
This configuration enables stable wave evolution, which realize a sequential framework that retrieves VOF field without adjusting boundary conditions. 

The paper is organized as follows. 
In Sec \ref{sec:methodology}, we provide a detailed description of the data assimilation method, including the problem statement. 
The wave reconstruction results are tested against the synthetic data from simulations in Sec \ref{sec:results}. 
Finally, we give the main conclusions of this work in Sec \ref{sec:conclusion}.

\section{Mathematical modeling and methodology}
\label{sec:methodology}
\subsection{Problem statement}

\begin{figure*}[h]
\centering
\includegraphics[width=0.9\textwidth]{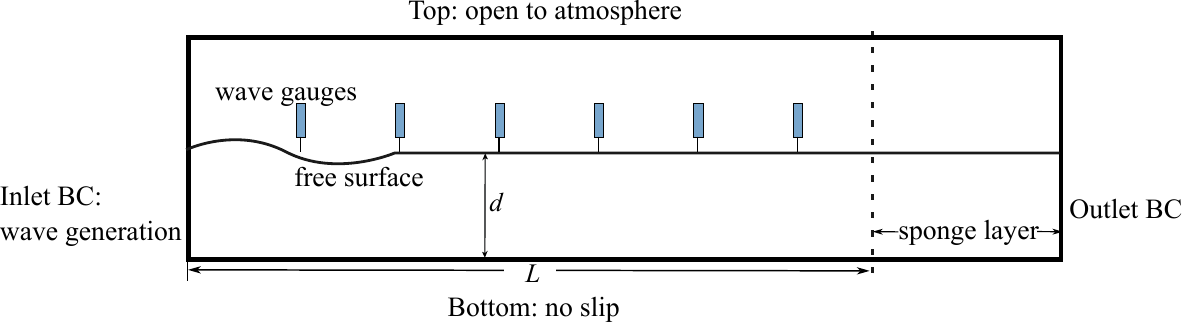}
\caption{Sketch of the numerical wave tank. The left boundary condition is the wave maker. Gauges are distributed without absorbing region. }
\label{fig:tank}
\end{figure*}


We consider an experimental water tank analogous to the two-dimensional numerical wave tank shown in Fig~\ref{fig:tank}.
The $x$ axis is the streamwise direction. The $z$ axis is the vertical direction. The origin is located at the still water level on the left boundary.
Measurements are collected at fixed locations along the tank. The sensors are distributed outside the absorbing region. 
These data are the free-surface elevation $\boldsymbol{y}(t)=\left(y_1(t), y_2(t), \cdots, y_{p_k}(t)\right)^T$, where, $p_k$ represents the total number of observations, e.g. the number of wave gauges. 
Besides, we assume that the observation error, from a statistical perspective, due to the precison of the measurement instruments is known.

As for the nonlinear dynamic model, the numerical wave tank can be simulated by any multiphase incompressible Navier-Stokes solvers, and the VOF field $\alpha(x, z, t)$ and the velocity fields $\boldsymbol{u}(x, z, t)=(u_x, u_z)^T$ are the primary outputs.
The objective of this study is to retrieve the analysis of model states $\alpha^a(x,z,t)$ given measurements $\boldsymbol{y}(t)$ in an optimal way, with the ultimate goal of reconstructing the water waves in a numerical wave tank and ensuring consistency with the actual wave propagation. 



Therefore, the governing equations are
\begin{align}
        \nabla \cdot \boldsymbol{u} =& 0, \\\
        \frac{\partial (\rho \boldsymbol{u})}{\partial t} + \nabla \cdot (\rho \boldsymbol{u} \otimes \boldsymbol{u}) =& -\nabla p + \nabla \cdot \left[ \mu \left( \nabla \boldsymbol{u} + (\nabla \boldsymbol{u})^T \right) \right] \nonumber + \rho \boldsymbol{g},\\\
        \frac{\partial \alpha}{\partial t} + \nabla \cdot (\alpha \boldsymbol{u}) =& 0.
\end{align}
Here, $\rho$ is the fluid density, $p$ is the pressure, $\mu$ is the dynamic viscosity, and $\boldsymbol{g}=(1, -g)^T$ with $g$ the gravity. 
The two phases considered in this study are air and water, and the density and viscosity are thus computed as  
\begin{align}
    \rho = \alpha \rho_w + (1-\alpha) \rho_a,\\
    \quad \mu = \alpha \mu_w + (1-\alpha) \mu_a.
\end{align}
$\rho_w$ and $\rho_a$ denote the densities of water and air, respectively, while $\mu_w$ and $\mu_a$ represent their corresponding dynamic viscosities.
$\alpha$ implicitly indicates the free surface, 1 for water, 0 for air, and the interface in between.
In this paper, OpenFOAM is employed to solve the nonlinear dynamic model. 
It is important to note that this choice is not a requirement for our methodology. 
For details on the numerical schemes, please refer to the paper by \citet{weller1998tensorial}.

\subsection{Wave reconstruction framework}

\begin{figure}\centering
\includegraphics[width=0.7\textwidth]{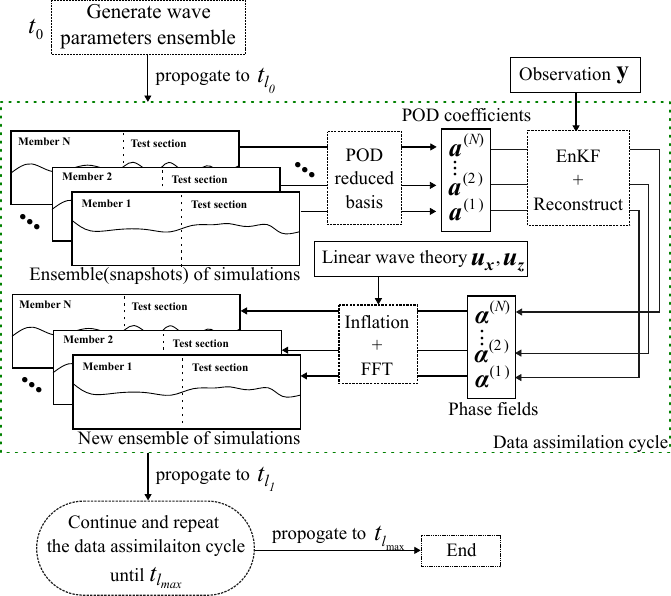}
\caption{Flow chart of the wave reconstruction framework.}
\label{fig:overall flowchart}
\end{figure}

The EnKF is the data assimilation method to incorporate measurements and the model output in this study.
Fig~\ref{fig:overall flowchart} shows the overall schematic procedure of the reconstruction framework.
The process begins with initial conditions $\boldsymbol{\alpha}(t_0)$, $\boldsymbol{u_x}(t_0)$, and $\boldsymbol{u_z}(t_0)$, the discretized versions of $\alpha(x,z,t_0)$, $u_x(x,z,t_0)$, and $u_z(x,z,t_0)$) as the baseline case.
Since the EnKF is a Monte Carlo method, an ensemble of simulations is required, the generation of which is detailed in Sec~\ref{sec:ensemble}.
Each ensemble member then evolves independently.

The first cycle of data assimilation starts at $t_{l_0}$.
POD is performed on the instantaneous snapshots of the ensemble members, specifically the fields $\boldsymbol{\alpha}^{(n)}(t_{l_0})$ where $n=1,2,\cdots, N$ with $N$ the ensemble size.
By default, the number of snapshots equals the ensemble size, although this can be configured differently.
This projects the fields into a reduced-order space, yielding the modal coefficients $\boldsymbol{a}^{(n),f}(t_{l_0})$.
The analysis step then corrects $\boldsymbol{a}^{(n),f}(t_{l_0})$ by integrating the corresponding ensemble member of measurements $\boldsymbol{y}^{(n)}(t_{l_0})$.
The updated state $\boldsymbol{a}^{(n),a}(t_{l_0})$ is projected back to the physical space and
the analysis of each ensemble member $\boldsymbol{\alpha}^{(n),a}(t_{l_0})$ is computed.
The POD process and the EnKF method are detailed in Sec~\ref{sec:SVD} and Sec~\ref{sec:EnKF} respectively.

To prevent ensemble collapse caused by underestimated uncertainty, inflation is applied after analysis. It maintains ensemble spread and helps the forecast covariance remain representative of the true variability. In this two phase flow problem, the core strategy is multiplicative inflation. Standard multiplicative inflation on the full state vector is not suitable in a VOF based wave tank. It can violate the boundedness and interface consistency of the volume fraction field. It can also produce nonphysical gas liquid distributions. For this reason, inflation is applied to the assimilated free surface. 
The implementation details are given in Sec~\ref{sec:inflation}.

This sequential data assimilation strategy will continue until the last cycle is finished at $t_{l_{max}}$.
To obtain a deterministic wave field retrieved at any given time, one can simply average the POD coefficients across all ensemble members and then project them back to physical space using the reduced basis corresponding to that specific moment.

\subsection{Generation of the ensemble}
\label{sec:ensemble}
The EnKF is a class of ensemble-based data assimilation methods that requires the construction of an ensemble of simulations to statistically represent the uncertainty in the initial state. 
Typically, a baseline is perturbed under a Gaussian distribution to generate an ensemble~\cite{da2018ensemble}.
This approach can be applied to the flow field $\boldsymbol{u}(x, z, t)$, as the solver can restore it to a divergence-free state within one time step.
However, as mentioned above, perturbations cannot be directly added to the phase field, since the VOF function $\alpha(x, z, t)$ is strictly bounded between 0 and 1.
Even if bounded perturbations are imposed, unphysical bubbles or droplets may emerge, which fails to preserve the physical constraints. 

Specifically, we prescribe perturbations to the core wave parameters—including wave height, period, and phase—for the inlet boundary condition at the beginning.
The ensemble of wave parameters is generated as
\begin{equation*}
    \beta^{(n)} = \beta^b+\epsilon,
    \label{eq:wave parameters}
\end{equation*}
where $\beta$ represents one of the wave parameter. 
$\beta^{(n)}$ is the special wave parameter of the $n$-th member. 
$\beta^b$ is the corresponding parameter of the baseline wave, and $\epsilon~ \sim \mathcal{N}(0, \sigma^2)$ a normally distributed perturbation with a user-defined standard deviation $\sigma$ to represent the uncertainty.
According to Fig~\ref{fig:overall flowchart}, by simulating each realization to time $t_{l_0}$, the initial statistical information of the flow fields used for data assimilation across the entire ensemble can be obtained.

\subsection{Free surface representation}
\label{sec:interface represent}




The VOF scaler field $\boldsymbol{\alpha}$ representing the phase of fluids is used to propagate the simulations.
The fluid volume fraction in each computational cell can be written as $\alpha(\boldsymbol{l})$ at a certain moment, where the cell location is $\boldsymbol{l}=(x, z)^T$. 
A value of $\alpha=0.5$ approximates the location of the interface.
The level-set function represents the interface as the zero contour of a signed distance function field $\boldsymbol{\phi}$, where $\phi>0$ in air, $\phi<0$ in water, and $\phi=0$ at the interface. It offers a smooth and differentiable representation, which is more suitable for POD to extract effective modes compared with VOF function.
We hence convert VOF data into level-set field using a signed distance function before entering into the POD and analysis step in one data assimilation cycle. 
For a given volume fraction $\alpha$, the corresponding level-set function $\phi$ for each cell is obtained via

\begin{equation}
\phi(\boldsymbol{l}) = 
\left\{ 
\begin{aligned}
&\underset{k \in \Gamma}{\min} \, ||\boldsymbol{l}-\boldsymbol{k}||, && \alpha(\boldsymbol{l}) < 0.5, \\
&-\underset{k \in \Gamma}{\min} \, ||\boldsymbol{l}-\boldsymbol{k}||, && \alpha(\boldsymbol{l}) > 0.5, \\
&0, && \alpha(\boldsymbol{l}) = 0.5.
\end{aligned}
\right.
\label{eq:levelset}
\end{equation}
$\Gamma$ is the free surface, and $\boldsymbol{k}$ is the point at this interface. 


In practice, the level-set method is used for modal decomposition and reconstruction as a temporary intermediate field, due to its more continuous structure. 
The VOF field, although more accurate in terms of mass conservation, may introduce higher noise due to its strongly discontinuous nature near the interface, and consequently requires more modes for effective representation. 
Some discussion will be provided in Sec~\ref{sec:results}.

\subsection{Modal decomposition via POD}
\label{sec:SVD}
The primary motivation for employing modal decomposition is to reduce the dimensionality of the flow field by projecting the high-dimensional data onto a modal space, thereby obtaining a compact set of modal coefficients that effectively represent the flow pattern. 
This dimensionality reduction significantly decreases the computational cost of the EnKF by operating in a reduced-order state space. 
Moreover, it mitigates the issues associated with directly assimilating high-dimensional state vectors, such as the introduction of non-smoothness, violation of physical constraints, and potential numerical instabilities or divergence. 
By capturing the dominant coherent structures of the flow, the modal representation ensures that the assimilation process remains as computational efficient and physics-constrained as possible.

The singular value decomposition (SVD) is performed to extract dominant modes for the level-set field $\boldsymbol{\phi}$ to resolve the free surface, as well as the velocity components $\boldsymbol{u_x}$ and $\boldsymbol{u_z}$ if necessary.
For example, at $t_{l_i}$ there is an ensemble of the phase fields $\boldsymbol{\alpha}^{(n),f}$ which can be converted into $\boldsymbol{\phi}^{(n),f}$.
A matrix of snapshots is constructed as
\begin{equation}
    \boldsymbol{S}_{\phi} = [\boldsymbol{\phi}^{(1),f}, \boldsymbol{\phi}^{(2),f}, \dots, \boldsymbol{\phi}^{(N),f}],
\end{equation}
where $\boldsymbol{\phi}^{(n),f}$ denotes the field of the $n$-th ensemble member at time $t_{l_i}$. 

SVD is then applied to the matrix as
\begin{equation}
    \boldsymbol{S}_{\phi} = \boldsymbol{U}_{\phi} \boldsymbol{\Sigma}_{\phi} \boldsymbol{V}_{\phi}^T,
\end{equation}
where $\boldsymbol{U}_{\phi}$ contains the spatial modes of the ensemble of $\boldsymbol{\phi}^{(n),f}$ as its columns, $\boldsymbol{\Sigma}_{\phi}$ is a diagonal matrix of singular values representing the relative energy of each mode, and $\boldsymbol{V}_{\phi}$ is the right singular vectors of $\boldsymbol{S}_{\phi}$. 
Then, a truncated modal matrix $\hat{\boldsymbol{U}}_{\phi}\in \mathbb{R}^{M \times r_{\phi}}$ is used to reduce the dimensionality containing the first $r_{\phi}$ leading modes, while $M$ is the number of cells of the computational domain and $M\gg r_{\phi}$.


The level-set field of each ensemble member is then projected onto the truncated modal space to form the forecast state $\boldsymbol{a}^{(n),f}$
\begin{equation}
    \boldsymbol{a}^{(n),f} = \hat{\boldsymbol{U}}_\phi^T \boldsymbol{\phi}^{(n),f}.
    \label{eq:coefficient vector}
\end{equation}

Each state $\boldsymbol{a}^{(n),f}$ is then put into the analysis step, where it is updated using the EnKF based on available observations. 
After the assimilation, we get the analysis state $\boldsymbol{a}^{(n),a}$.
The updated level-set field can accordingly obtained through
\begin{equation}
\boldsymbol{\phi}^{(n),a} = \hat{\boldsymbol{U}}_\phi \, \boldsymbol{a}^{(n),a}.
\end{equation}
An inverse operator of Eq(\ref{eq:levelset}) is implemented to convert $\boldsymbol{\phi}^{(n),a}$ into the reconstructed phase field $\boldsymbol{\alpha}^{(n),a}$ at $t_{l_i}$.



\subsection{EnKF}
\label{sec:EnKF}

The EnKF is a Monte Carlo–based Bayesian filtering method, well-suited for high-dimensional and nonlinear systems. 
Its core idea is to approximate the covariance structure of the system state using a finite ensemble of samples, and to update these samples based on observations to obtain the analysis state.

As mentioned above, each member of the forecast state ensemble is denoted as $\boldsymbol{a}^{(n),f}$. 
The ensemble mean is given by
\begin{equation}
    \boldsymbol{\bar a}^f = \frac{1}{N} \sum_{n=1}^{N} \boldsymbol{a}^{(n),f}.
\end{equation}

The observation $\boldsymbol{y}$ is measured at $p_k$ positions. 
To simulate realistic measurement errors and satisfy EnKF assumptions, Gaussian noise is added to $\boldsymbol{y}$ to generate an ensemble of perturbed observations.
Each member is $\boldsymbol{y}^{(n)} = \boldsymbol{y} + \boldsymbol{\epsilon_o}$, and $\boldsymbol{\epsilon_o} \sim \mathcal{N}(0, \boldsymbol{R})$, with $\boldsymbol{R}$ being the observation error covariance matrix. 
In this paper, we simplify $\boldsymbol{R}=\sigma_o^2 \boldsymbol{I}$, where \(\sigma_o^2\) is the constant variance of the observation noise and \(\boldsymbol{I}\) is the identity matrix of appropriate dimensions.

During the EnKF analysis step, each forecast sample is updated by incorporating the corresponding perturbed observation. 
The update formula for the $n$-th ensemble member is
\begin{equation}
    \boldsymbol{a}^{(n),a} = \boldsymbol{a}^{(n),f} + \boldsymbol{K} \left( \boldsymbol{y}^{(n)} - \mathcal{H}(\boldsymbol{a}^{(n),f}) \right),
\end{equation}
where $\boldsymbol{a}^{(n),a}$ is the analysis state, $\boldsymbol{K}$ is the Kalman gain matrix, and $\mathcal{H}(\cdot)$ denotes the observation operator that maps the modal coefficients to the observation space. 
The detailed derivation of the observation operator can be found in the \ref{sec:obervation operator}.

The Kalman gain $\boldsymbol{K}$ is given by
\begin{equation}
    \boldsymbol{K} = \boldsymbol{P}^f \boldsymbol{H}^T \left( \boldsymbol{H} \boldsymbol{P}^f \boldsymbol{H}^T + \boldsymbol{R} \right)^{-1},
\end{equation}
where $\boldsymbol{P}^f$ is the forecast error covariance
\begin{equation}
    \bm P^f = \frac{1}{N-1}\sum_{n=1}^N\left(\boldsymbol{a}^{(n),f}- \boldsymbol{\bar a}^{f}\right)\left(\boldsymbol{a}^{(n),f}-\boldsymbol{\bar a}^{f}\right)^T,
\end{equation}
and
\begin{eqnarray}
    \bm P^f\bm H^T &=& \frac{1}{N-1}\sum_{n=1}^N\left(\boldsymbol{a}^{(n),f}-\boldsymbol{\bar a}^{f}\right)\left(\mathcal{H}(\boldsymbol{a}^{(n),f})-\overline{\mathcal{H}(\boldsymbol{a}^{f})}\right)^T,\\
    \bm H\bm P^f\bm H^T &=& \frac{1}{N-1}\sum_{n=1}^N\left(\mathcal{H}(\boldsymbol{a}^{(n),f})-\overline{\mathcal{H}(\boldsymbol{ a}^{f})}\right)\left(\mathcal{H}(\boldsymbol{a}^{(n),f})-\overline{\mathcal{H}(\boldsymbol{ a}^{f})}\right)^T.
\end{eqnarray} 

The ensemble mean of the analysis states serves as the optimal estimate used for the deterministic wave field reconstruction
\begin{equation}
    \boldsymbol{\bar a}^a = \frac{1}{N} \sum_{n=1}^{N} \boldsymbol{a}^{(n),a}.
\end{equation}

\subsection{Inflation and consistency}
\label{sec:inflation}

Inflation is used to mitigate sampling errors and ensemble collapse in the EnKF. Its implementation is difficult in a VOF based numerical wave tank. The VOF function $\alpha(x,z,t)$ is bounded between 0 and 1. A direct inflation of $\boldsymbol{\alpha}$ can violate boundedness and can smear the interface. Inflation must also preserve the coupling between the free surface and the velocity field. If the free surface is inflated without a consistent velocity adjustment, the next forecast can contain strong spurious transients.

At an assimilation instant $t_{l_0}$, the analysis ensemble $\boldsymbol{\alpha}^{(n),a}(t_{l_0})$ is first converted to a surface elevation field $\eta^{(n)}(x,t_{l_0})$. Inflation is not applied pointwise to $\eta^{(n)}$. We apply the fast Fourier transform (FFT) along $x$ axis. Each component is represented by $(A_i,k_i,\omega_i,\varphi_i)$, where $A_i$ is the amplitude, $k_i$ is the wavenumber, $\omega_i$ is the angular frequency, $\varphi_i$ is the phase. Inflation is applied to the amplitudes $A_i$.

\begin{equation}
A_i^{(n)}=\bar A_i+\varrho\left(A_i^{(n)}-\bar A_i\right),
\end{equation}
where $\bar{A}_i$ is the average amplitude of the $i$-th component of the ensemble.

The inflated surface elevation $\eta^{(n)}(x,t_{l_0})$ is then reconstructed by the inverse transform. A new ensemble $\boldsymbol{\alpha}^{(n)}(t_{l_0})$ is rebuilt from the inflated $\eta^{(n)}(x,t_{l_0})$. This step enforces the boundedness of $\alpha$ and restores an interface consistent with the inflated free surface.

The velocity fields corresponding to the inflated wave components are computed using potential wave theory.
The velocity components in the $x$ and $z$ directions are calculated as follows

\begin{align}
  u_{wx}(x, z, t_{l_0}) = \sum_i^m \frac{A_i g k_i}{\omega_i} \frac{\cosh(k_i(z + d))}{\cosh(k_i d)} \sin(k_i x + \varphi_i), \\
  u_{wz}(x, z, t_{l_0}) = \sum_i^m \frac{A_i g k_i}{\omega_i} \frac{\sinh(k_i(z + d))}{\cosh(k_i d)} \sin(k_i x + \varphi_i), 
\end{align}
where $d$ is the water depth.

The new ensemble of velocity fields is then reconstructed as
\begin{align}
    \boldsymbol{u_x}^{(n)}(t_{l_0}) = u_{wx}^{(n)}(x, z, t_{l_0}) \boldsymbol{\alpha}^{(n)}(t_{l_0}), \\
    \boldsymbol{u_z}^{(n)}(t_{l_0}) = u_{wz}^{(n)}(x, z, t_{l_0}) \boldsymbol{\alpha}^{(n)}(t_{l_0}).
\end{align}
These velocity fields are used to propagate the waves in the VOF-based numerical wave tank until the next assimilation cycle at time $t_{l_1}$.

Although multiplicative inflation is commonly tuned with a coefficient in the interval $[1.0, 1.2]$ \cite{Bai2011}, the standard tuning guidance cannot be transferred without verification. This is because the analysis spread that controls the filter update is modified through the reconstructed free surface, rather than through direct inflation of the state vector.
A consistency analysis is therefore required. The present inflation changes the ensemble variability in the observation-related subspace by amplifying spectral amplitudes of the assimilated free surface. It also relies on an auxiliary velocity field after inflation. 
Both design alter the relationship between ensemble spread and the actual forecast error. Without a consistency check, the inflated ensemble can become under-dispersive or over-dispersive. Either case degrades the sequential EnKF performance and can bias the reconstruction. 

To evaluate the consistency of the assimilation process, we use the Normalized Estimation Error Squared (NEES) as a diagnostic tool\cite{barshalom2001estimation, ivanov2014consistency}. This study is based on a twin numerical experiment, where the true state is known. The NEES is computed at each assimilation time step to assess whether the analysis uncertainty is statistically consistent with the actual estimation error. 

At each assimilation time step $t_{l_i}$, the analysis ensemble $\{\boldsymbol{a}^{(n),a}(t_{l_i})\}_{n=1}^{N}$ yields the analysis mean $\boldsymbol{\bar a}^a(t_{l_i})$ and the analysis error covariance
\begin{equation}
    \boldsymbol{P}^a(t_{l_i}) = \frac{1}{N-1} \sum_{n=1}^{N} \left(\boldsymbol{a}^{(n),a}(t_{l_i}) - \boldsymbol{\bar a}^{a}(t_{l_i})\right) \left(\boldsymbol{a}^{(n),a}(t_{l_i}) - \boldsymbol{\bar a}^{a}(t_{l_i})\right)^T.
\end{equation}

The true coefficient vector is obtained by projecting the true level-set field onto the same truncated basis
\begin{equation}
    \boldsymbol{a}^{t}(t_{l_i}) = \hat{\boldsymbol{U}}_{\phi}^{T} \boldsymbol{\phi}^{t}(t_{l_i}),
\end{equation}
where $\boldsymbol{\phi}^{t}(t_{l_i})$ denotes the true level-set field. The NEES is then calculated as
\begin{equation}
    \mathrm{NEES}(t_{l_i}) = \left(\boldsymbol{a}^{t}(t_{l_i}) - \boldsymbol{\bar a}^{a}(t_{l_i})\right)^T \left(\boldsymbol{P}^{a}(t_{l_i})\right)^{-1} \left(\boldsymbol{a}^{t}(t_{l_i}) - \boldsymbol{\bar a}^{a}(t_{l_i})\right).
\end{equation}
Under Gaussian assumptions, a consistent estimator yields NEES values that follow a chi-square distribution with $r_{\phi}$ degrees of freedom, where $r_{\phi}$ is the dimension of the modal coefficient vector $\boldsymbol{a}$. The normalized NEES, defined as $\mathrm{NEES}(t_{l_i})/r_{\phi}$, is reported to facilitate comparison across different truncation settings.

By monitoring the NEES values, the optimal inflation factor $\varrho$ can be selected to ensure the best filter consistency.

\section{RESULTS AND DISCUSSION}
\label{sec:results}

To evaluate the effectiveness and applicability of this method under various wave conditions, we consider three representative conditions in this study: regular waves, irregular waves, and plunging waves. These three wave conditions represent a progression in flow complexity. Concretely, fifth-order Stokes waves are used to represent regular waves. JONSWAP-spectrum waves represent irregular wave conditions. And in the plunging wave case, a cosine wave is made to climb a slope, leading to the formation of a plunging breaker.  

In the following sections, we present the reconstruction results for each wave case. This analysis provides insights into the performance and limitations of the framework under challenging conditions.

\subsection{Regular Wave Reconstruction}
\label{sec:stokes V}

This subsection examines the instantaneous assimilation performance for a regular fifth-order Stokes wave. This wave condition is characterized by a single dominant frequency and a stable phase relation. The free surface and the associated velocity field are smooth and repeatable. The reconstruction difficulty is therefore lower than that of irregular waves. The purpose of this test is to validate the ability of the proposed framework to correct a simple wave field at a given assimilation time, including both the free surface and the velocity components, before moving to more demanding non-periodic cases. 

The twin experiment is configured as follows. The true wave has a wave height $H^t$ of 0.15~m. The baseline wave has a wave height $ H^b$ of 0.1~m. The wave period $T$ is identical in the two cases and equals 2~s. The water depth $d$ is 0.60~m. Observations are free surface elevations measured at 10 locations that are uniformly distributed along the computational domain. The observation noise standard deviation $\epsilon_o$ is 0.002 m.

\begin{figure}
\centering
\includegraphics[width=0.7\textwidth]{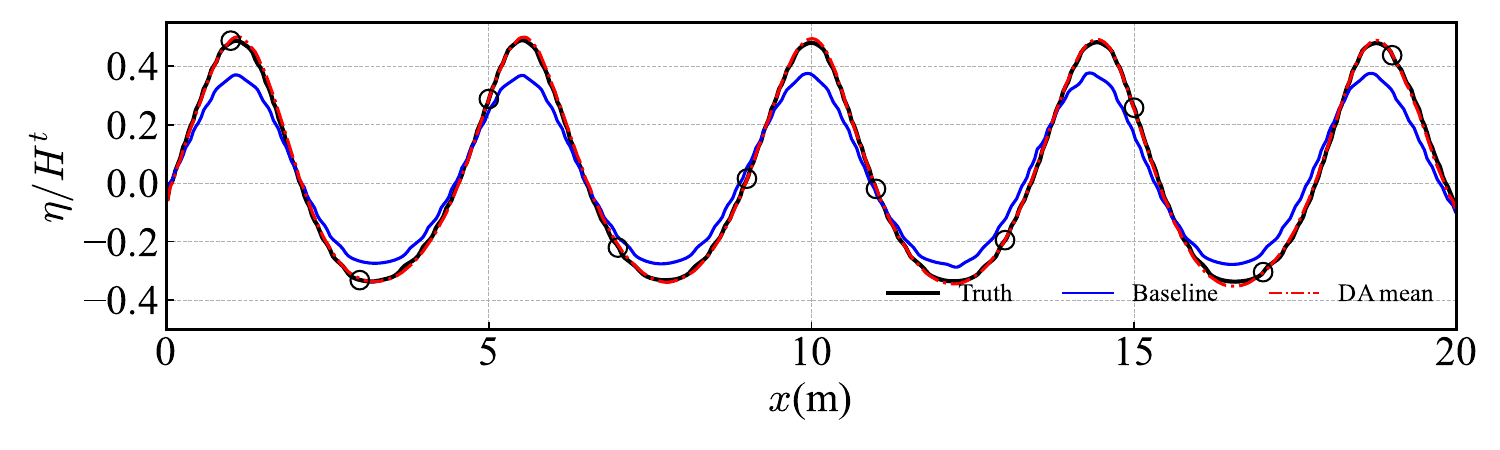}
\caption{Surface elevations $\eta^a(x)$, $\eta^b(x)$, and $\eta^t(x)$ at given assimilation instant($t_{l_0}=20$~s), cycles represent elevation observations.}
\label{fig:stokes height_20s}
\end{figure}

The assimilation process is triggered at $t_{l_0}=20$~s.
At this moment, the wave train has stably propagated through the computational domain.
We first evaluate the correction of the free surface elevation. Here, we adopt level-set to represent phase field.
Fig~\ref{fig:stokes height_20s} compares the free surface profiles of the truth, the baseline, and the DA mean. 
The vertical axis represents the elevation normalized by the true wave height $H^t$.
The baseline simulation captures the wave phase accurately.
This is expected as the wave period $T$ is identical to the true state.
However, a significant discrepancy exists in the wave amplitude.
The data assimilation effectively corrects this amplitude error.
The reconstructed surface profile, indicated by the red dash-dotted line, shows excellent agreement with the true state (black line).
This result confirms that the EnKF properly updates the level-set field coefficients to match the measured elevations.

\begin{figure}[htbp]
\centering

\begin{subfigure}[b]{0.7\textwidth}
    \centering
    \includegraphics[width=\textwidth]{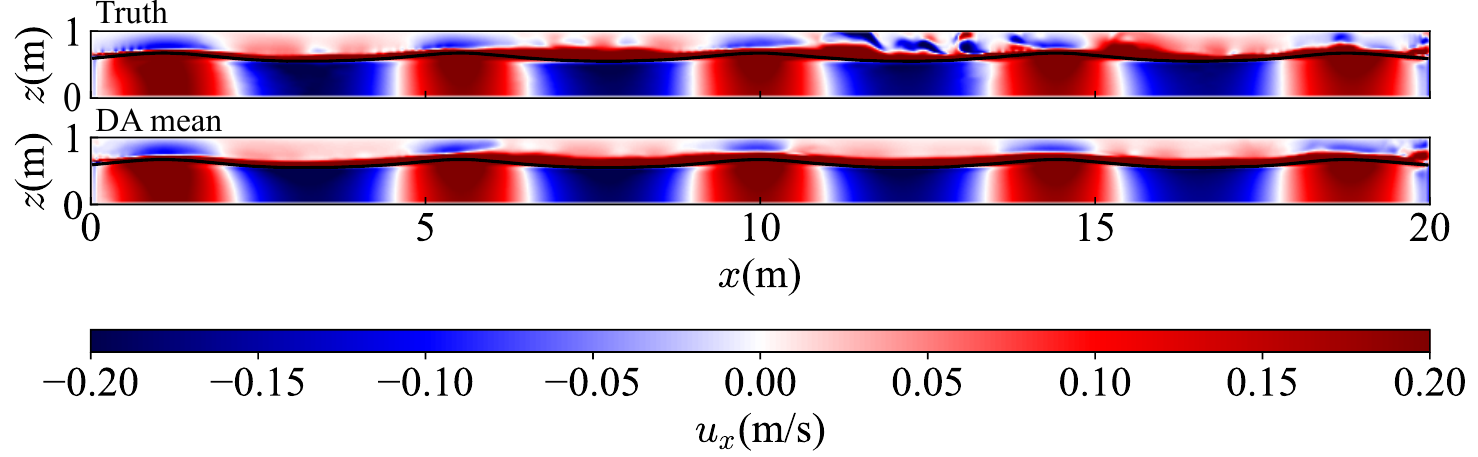}
    \caption{${u}_x$ comparison}
    \label{fig:stokes ux}
\end{subfigure} 

\begin{subfigure}[b]{0.7\textwidth}
    \centering
    \includegraphics[width=\textwidth]{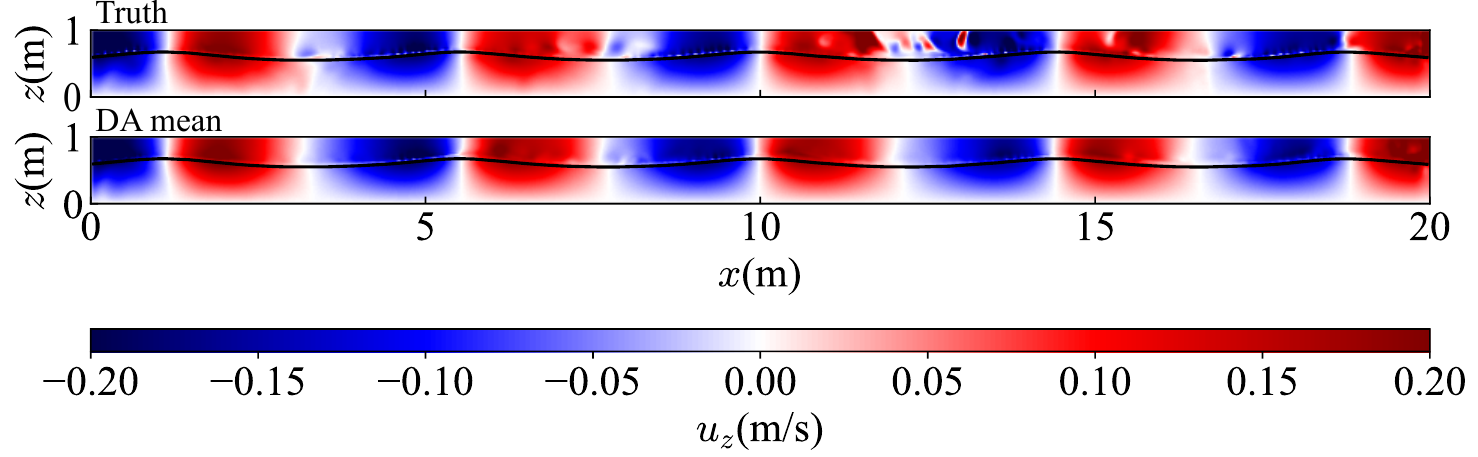}
    \caption{${u}_z$ comparison}
    \label{fig:stokes uz}
\end{subfigure}

\caption{Comparison of reconstructed and true velocity fields at the assimilation instant($t_{l_0}=20$~s).}
\label{fig:stokes velocity}
\end{figure}

In this test, the velocity components are included in the state vector and assimilated with the level-set state.
Fig~\ref{fig:stokes ux} and \ref{fig:stokes uz} present the contours of the horizontal velocity component $\boldsymbol{u}_x$ and the vertical velocity component $\boldsymbol{u}_z$, respectively.
The top panels display the true state, while the bottom panels display the assimilated state.
The assimilated velocity fields exhibit the characteristic motion of Stokes waves.
In Fig~\ref{fig:stokes ux}, the horizontal velocity $\boldsymbol{u}_x$ is positive under the wave crests and negative under the troughs.
The spatial distribution in the DA result matches the true pattern.
Similarly, Fig~\ref{fig:stokes uz} shows the phase shift in the vertical velocity $\boldsymbol{u}_z$.
The vertical velocity is upward on the wave front and downward on the wave back.

It is worth noting that the true velocity fields in the air phase exhibit visible disturbances, particularly in the region above the interface.
These fluctuations are attributed to the large density ratio between water and air in the VOF solver, where the lighter phase is highly sensitive to pressure gradients, leading to spurious parasitic currents.
Despite these disturbances in the air, the wave profile itself remains intact and physically consistent.
In contrast, the DA reconstructed fields do not reproduce these non-physical fluctuations.
The air phase in the assimilated results is remarkably smooth.
This filtering effect is attributed to the modal truncation inherent in the POD representation.
The parasitic currents manifest as high-frequency and low-energy noise.
Such features generally correspond to higher-order modes that are excluded from the truncated basis, or the EnKF analysis step would not correct these delicate perturbations.
Consequently, the reconstruction effectively filters out the numerical noise present in the raw CFD data while preserving the essential flow physics. 

The determination of the dimension of the truncated subspace $r$ is a critical step in the POD-EnKF framework.
A common criterion is to retain enough modes to capture a specific percentage of the total fluctuating energy, defined by the cumulative variance.
In this study, a threshold of $99\%$ is adopted to ensure that the dominant flow structures and the wave profile are accurately preserved while filtering out low-energy noise.

\begin{figure*}
    \centering
    \begin{subfigure}[b]{0.32\textwidth}
        \centering
        \includegraphics[width=\textwidth]{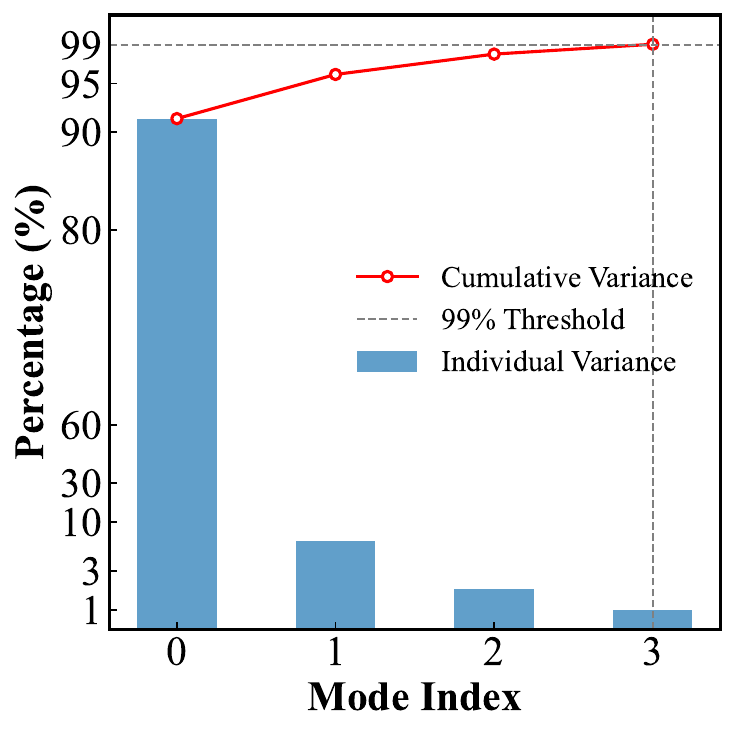}
        \caption{$\hat{U}_{\phi}$}
        \label{fig:phi_eig}
    \end{subfigure}
    \hfill
    \begin{subfigure}[b]{0.32\textwidth}
        \centering
        \includegraphics[width=\textwidth]{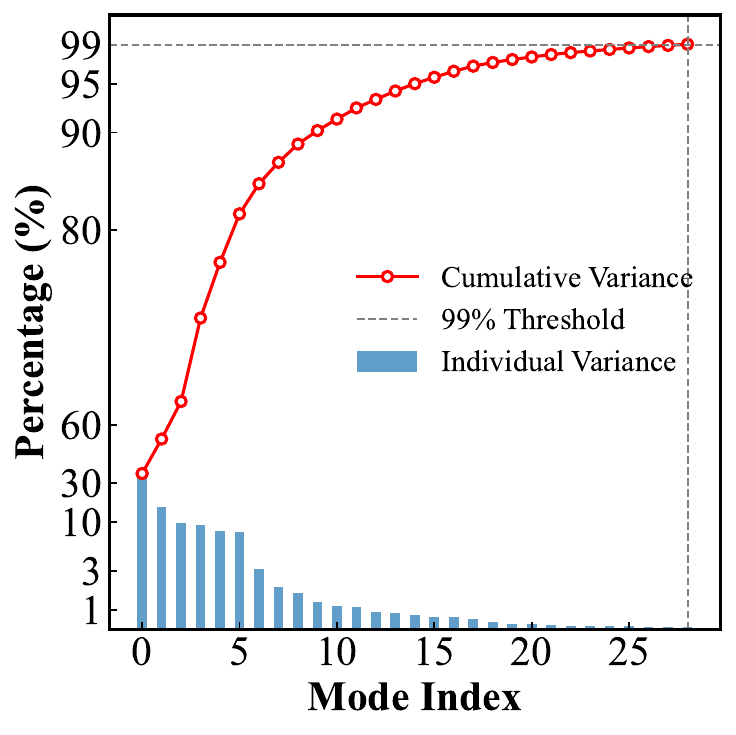}
        \caption{$\hat{U}_{u_x}$}
        \label{fig:Ux_eig}
    \end{subfigure}
    \hfill
    \begin{subfigure}[b]{0.32\textwidth}
        \centering
        \includegraphics[width=\textwidth]{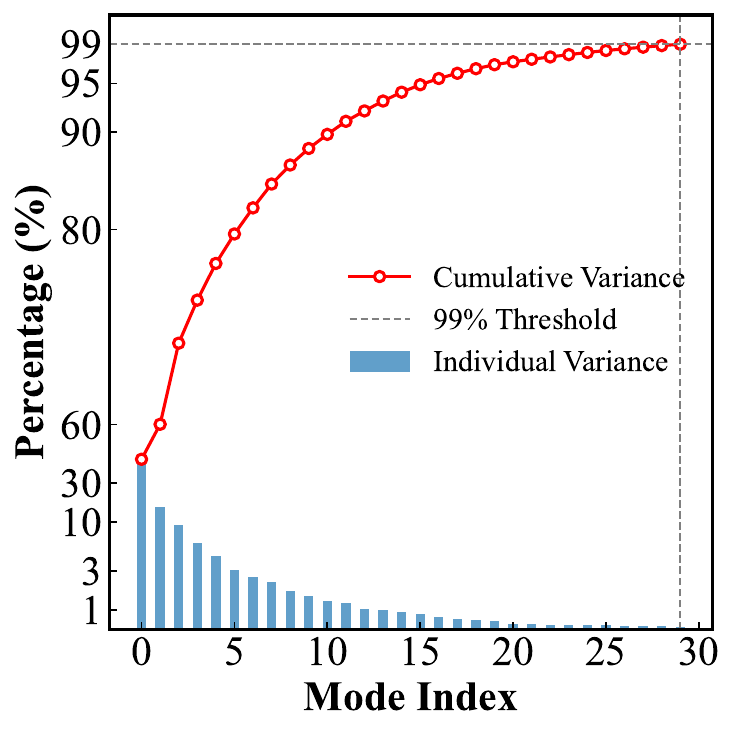}
        \caption{$\hat{U}_{u_z}$}
        \label{fig:Uz_eig}
    \end{subfigure}
    \caption{Energy distribution of truncated POD modes for the regular wave case. The blue bars denote individual variance, and the red line shows cumulative variance. The dashed line marks the truncation point for $99\%$ energy.}
    \label{fig:stokes eigenvalue}
\end{figure*}

Fig~\ref{fig:stokes eigenvalue} presents the Pareto charts of the singular values for the level-set $\boldsymbol{\phi}$, the horizontal velocity $\boldsymbol{u}_x$, and the vertical velocity $\boldsymbol{u}_z$.
The blue bars represent the individual variance of each mode, while the red line indicates the cumulative variance.
Different characteristics are observed between the phase field and the velocity fields.
As shown in Fig~\ref{fig:phi_eig}, the energy of the level-set field is highly concentrated in the leading modes.
The first mode (mode 0) alone accounts for over $90\%$ of the total variance.
The cumulative variance rapidly surpasses the $99\%$ threshold at mode 3.
Consequently, only the first 4 modes are needed to reconstruct the regular wave interface with high fidelity.
This high compressibility is attributed to the smoothness of the signed distance function used to represent the interface.
In contrast, the velocity fields $\boldsymbol{u}_x$ and $\boldsymbol{u}_z$ display a more slowly decaying energy spectrum, as illustrated in Fig~\ref{fig:Ux_eig} and \ref{fig:Uz_eig}.
The first mode contributes less than $40\%$ and $50\%$ to the total energy for $\boldsymbol{u}_x$ and $\boldsymbol{u}_z$, respectively.
The energy is distributed across a wider range of spatial modes.
To reach the $99\%$ energy criterion, significantly more modes are required—specifically, 29 modes for $\boldsymbol{u}_x$ and 30 modes for $\boldsymbol{u}_z$.
This implies that the velocity field involves more complex spatial features than the level-set field.
Nevertheless, the truncated basis effectively reduces the degrees of freedom from the order of mesh size ($M=4.8 \times 10^{4}$) to approximately 30, greatly facilitating the ensemble-based assimilation.

\begin{figure*}
    \centering
    \begin{subfigure}[b]{0.32\textwidth}
        \centering
        \includegraphics[width=\textwidth]{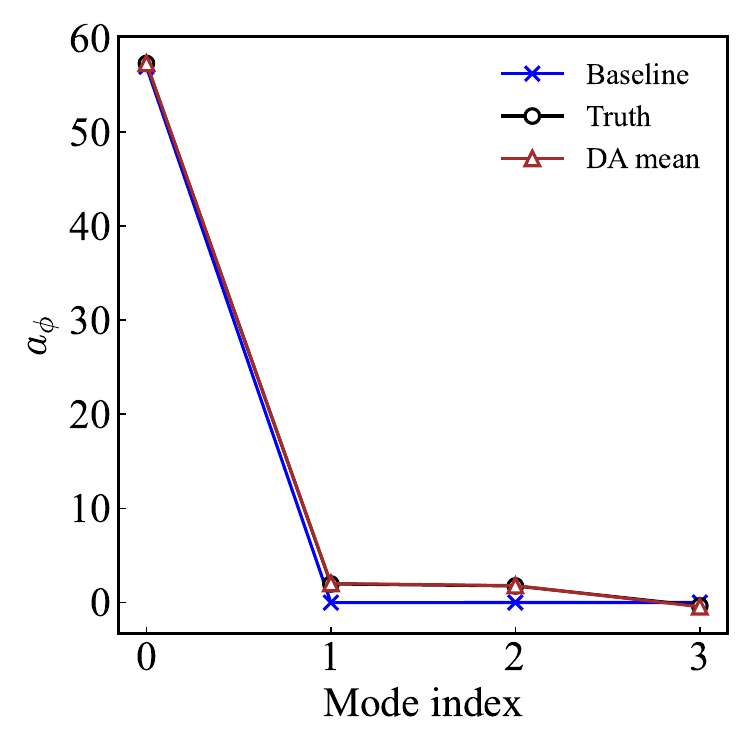}
        \caption{$\boldsymbol{a}_{\phi}$}
        \label{fig:phi_coeff}
    \end{subfigure}
    \hfill
    \begin{subfigure}[b]{0.32\textwidth}
        \centering
        \includegraphics[width=\textwidth]{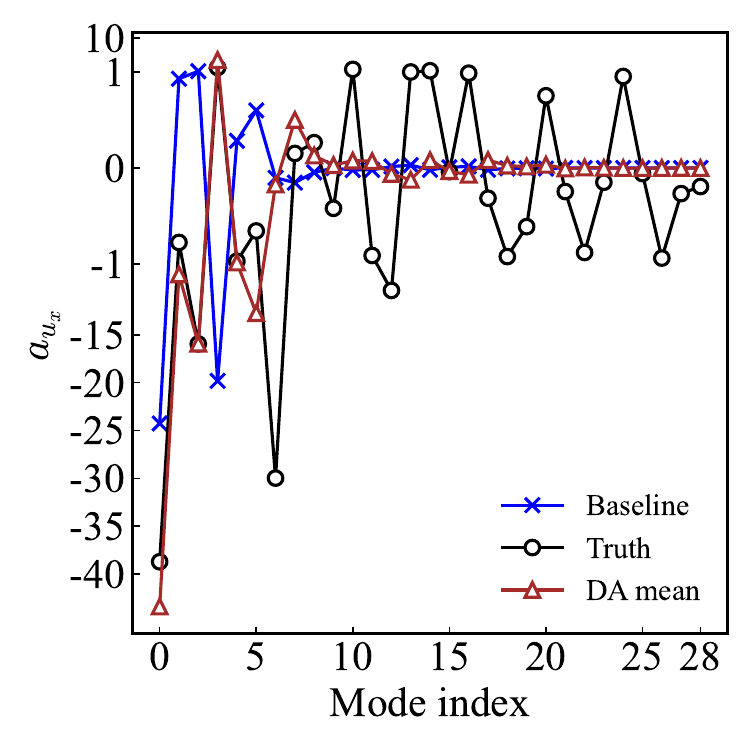}
        \caption{$\boldsymbol{a}_{u_x}$}
        \label{fig:Ux_coeff}
    \end{subfigure}
    \hfill
    \begin{subfigure}[b]{0.32\textwidth}
        \centering
        \includegraphics[width=\textwidth]{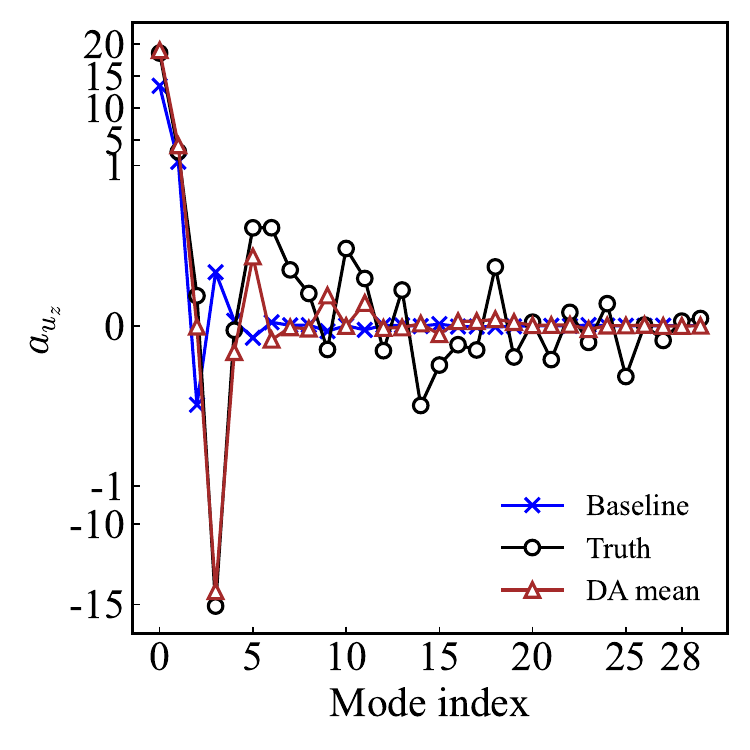}
        \caption{$\boldsymbol{a}_{u_z}$}
        \label{fig:Uz_coeff}
    \end{subfigure}
    \caption{Comparison of modal coefficients before and after assimilation. The baseline is represented as blue crosses, the DA mean is represented as red triangles, and the truth is represented as black circles.}
    \label{fig:stokes coefficients}
\end{figure*}

Following the determination of the truncated subspace, we check the update performance of the state vectors within this reduced-order space.
Fig~\ref{fig:stokes coefficients} compares the modal coefficients of the baseline, the truth, and the DA analysis mean for the level-set and velocity components.
This spectral comparison reveals how the EnKF adjusts the weights of different flow structures to achieve reconstruction.

For the level-set, shown in Fig~\ref{fig:phi_coeff}, the correction is straightforward.
Consistent with the energy analysis, the zeroth mode dominates the magnitude.
For the higher-order modes (modes 1-3), although their magnitudes are small, the DA still effectively captures the subtle variations present in the truth, ensuring the precise reconstruction of the interface shape.

The velocity coefficients $\boldsymbol{a}_{u_x}$ and $\boldsymbol{a}_{u_z}$, displayed in Fig~\ref{fig:Ux_coeff} and \ref{fig:Uz_coeff}.
In the low-frequency range (e.g., mode indices 0-5), which contains the majority of the kinetic energy, the baseline deviates considerably from the truth.
The assimilation roughly corrects these leading coefficients, restoring the correct wave orbital motions.
However, a significant divergence appears in the higher-order modes.
The true coefficients exhibit persistent fluctuations.
Crucially, the DA analysis mean does not reproduce these fluctuations; instead, its coefficients decay smoothly to zero.
This behavior is attributed to the statistical nature of the EnKF.
The parasitic currents appearing in the ensemble members are typically random and incoherent.
Consequently, they do not form a consistent correlation structure within the forecast error covariance matrix.
The Kalman gain, therefore, filters out these uncorrelated non-physical features.

The results presented above demonstrate that the proposed EnKF-based framework effectively reconstructs the regular wave field at a specific instant.
By assimilating wave elevation data from sparse gauges, the method successfully corrects the amplitude discrepancy in the baseline simulation.
The reconstructed free surface matches the truth with high accuracy.
Furthermore, the velocity field is retrieved simultaneously through the covariance inherent in the ensemble.
This validates the feasibility of using a truncated POD subspace for state estimation in VOF-based wave tanks.

\begin{figure}[htbp]
  \centering
  \begin{minipage}[t]{0.49\textwidth}
    \centering
    \includegraphics[width=0.7\linewidth]{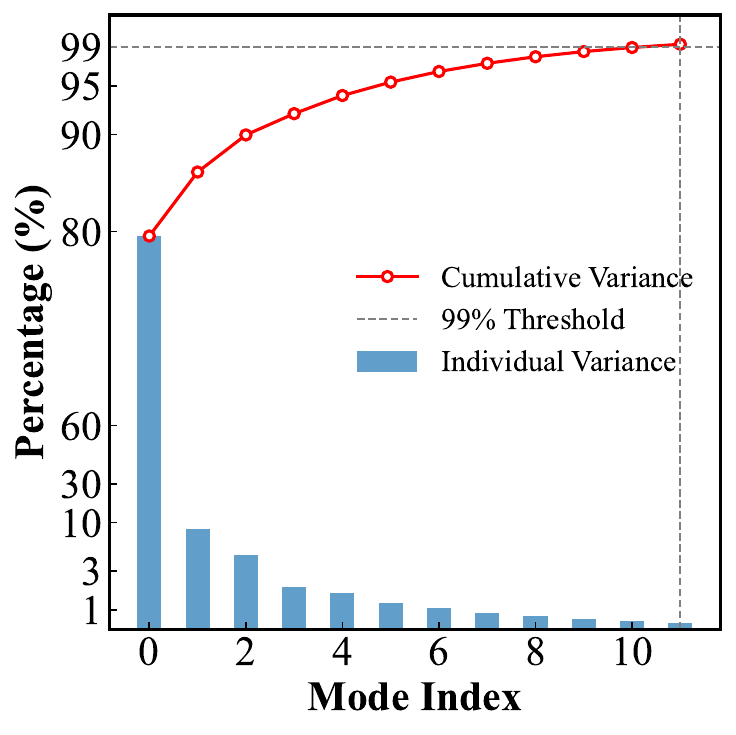}
    \caption{Energy distribution of POD modes for the VOF field in fifth-order Stokes waves.}
    \label{fig:vof_eig}
  \end{minipage}
  \hfill
  \begin{minipage}[t]{0.49\textwidth}
    \centering
    \includegraphics[width=0.7\linewidth]{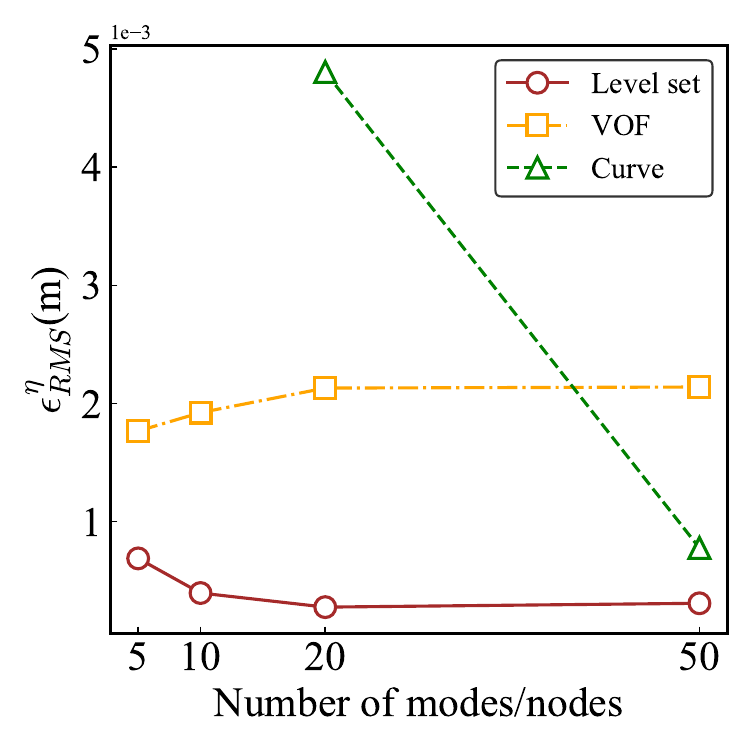}
    \caption{Comparison of free surface reconstruction error among different interface representation methods.}
    \label{fig:representation_compare}
  \end{minipage}
\end{figure}

We now examine how the choice of representation for the free surface affects the reconstruction accuracy.
Two representations are compared: the VOF field $\boldsymbol{\alpha}$ and the level-set field $\boldsymbol{\phi}$.
The reconstruction quality depends heavily on the compactness of the modal basis, which is determined by the smoothness of the field.
Fig~\ref{fig:vof_eig} shows the energy distribution of the POD modes for the VOF field.
In contrast to the level-set field shown in Fig~\ref{fig:phi_eig}, the VOF energy spectrum decays slowly.
The first mode accounts for approximately $80\%$ of the energy, compared to over $90\%$ for the level-set method.
To reach the $99\%$ cumulative energy threshold, the VOF representation requires 11 modes, whereas the level-set representation requires only 4.
This inefficiency arises from the binary nature of the VOF method (0 or 1).
The discontinuity at the interface introduces high-frequency spectral components that are difficult to capture with a low-rank linear basis.


To quantify the reconstruction performance, we compare the Root Mean Square (RMS) error of the free surface elevation across different representation methods.
The elevation RMS is computed as 

\begin{align}
    e_{\eta}(x) = \eta(x)-\eta^{t}(x),\\
    \epsilon_{RMS}^{\eta}=\sqrt{\frac{1}{L}\int_{0}^{L}\left(e_{\eta}(x)\right)^2 \, \mathrm{d}x },
\end{align}
where $\eta(x)$ denotes the reconstructed free-surface elevation, and $\eta^{t}(x)$ is the corresponding true elevation. And $e_{\eta}$ is the absolute error of the elevation. With uniform spatial sampling $\Delta x$ and $Q=L/\Delta x$, the integral is approximated by
\begin{equation}
\epsilon_{RMS}^{\eta}\approx \sqrt{\frac{1}{Q}\sum_{q=1}^{Q}\left(e_{\eta}(x_q)\right)^2 } .
\end{equation}

Fig~\ref{fig:representation_compare} presents the RMS error of the reconstructed evolution.
In this comparison, a parametric curve representation is introduced as a baseline reference.
This method discretizes the free surface using a set of nodes uniformly distributed along the wave propagation direction, with the profile reconstructed via interpolation.
Since the curve representation directly tracks the interface geometry with inherent smoothness, it serves as a high-precision benchmark for the regular wave case.
The VOF method exhibits the largest error, significantly higher than the other two approaches when adopting 50 modes.
In contrast, the level-set method demonstrates superior accuracy.
Remarkably, it achieves the lowest error magnitude, matching and even slightly outperforming the curve-based benchmark.
This result indicates that converting the VOF field to a level-set field effectively improves the accuracy of reconstruction.

Based on these findings, the level-set representation is selected as the optimal choice for the proposed framework.
While the curve method performs well for this regular wave, it is strictly limited to single-valued surfaces.
The level-set method offers a balanced solution: it attains the high reconstruction accuracy of the parametric curve while retaining the topological flexibility of the VOF method to describe complex, multi-valued interfaces.


To assess the robustness of the proposed framework, we investigate the influence of the observation noise magnitude on the reconstruction accuracy.
The observation noise standard deviation $\epsilon_o$ is varied from $10^{-4}$~m to $10^{-1}$~m, covering a range from high-precision laboratory measurements to potentially noisy field data.
The reconstruction quality is evaluated using RMS error.
The accuracy of the velocity field is assessed globally across the entire computational domain.
The RMS error for a velocity component $u$ (representing either $u_x$ or $u_z$) is defined as:
\begin{equation}
    \epsilon_{RMS}^u = \sqrt{ \frac{1}{M} \sum_{i=1}^{M} \left( u_i - u_i^t \right)^2 },
\end{equation}
where $M$ denotes the total number of cells in the domain, and $u_i$ and $u_i^t$ correspond to the analyzed and true velocity values at the $i$-th grid, respectively.

\begin{figure}
\centering
\includegraphics[width=0.33\textwidth]{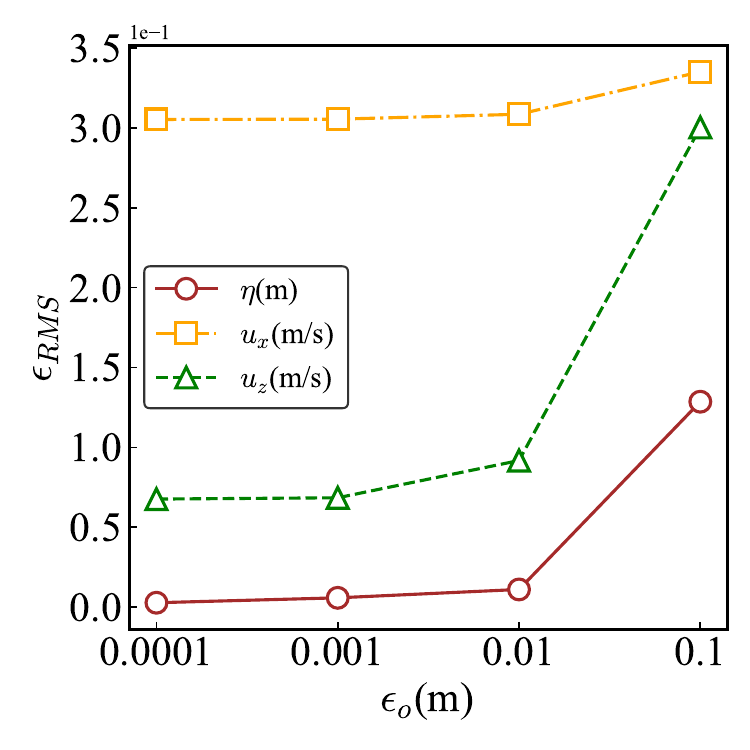}
\caption{Reconstruction RMS errors for free surface elevation, horizontal velocity, and vertical velocity for different observation noise standard deviations.}
\label{fig:obsstd}
\end{figure}

Fig~\ref{fig:obsstd} presents the variations of $\epsilon_{RMS}$ for the free surface elevation, the horizontal velocity, and the vertical velocity with respect to the observation noise level.
The results indicate a clear correlation between the measurement precision and the reconstruction accuracy.
In the regime of low observation noise ($\epsilon_o < 10^{-3}$~m), the errors for all quantities remain relatively low and stable.
In this range, the reconstruction performance is primarily limited by the modal truncation error and the ensemble spread rather than the measurement uncertainty.
As $\epsilon_o$ increases beyond $10^{-3}$~m, a gradual increase in the reconstruction error is observed.
Notably, the errors in the unobserved velocity components ($u_x$ and $u_z$) follow the same trend as the observed free surface $\eta$.
This synchronous behavior confirms that the ensemble-based covariance matrix effectively propagates the correction from the surface observations to the flow fields, even under varying noise conditions.

\subsection{Irregular Wave Reconstruction}
This subsection extends the validation to the more complex scenario of irregular waves.
The primary objective is to verify the sequential assimilation capability of the proposed EnKF framework for the phase-resolved reconstruction of a stochastic wave train in a VOF-based numerical wave tank.

A fundamental challenge in phase-resolved reconstruction lies in the stochastic nature of the boundary conditions.
For irregular seas, the wave field is not deterministic. While the target spectrum describes the energy distribution, the specific realization of the wave train is governed by random phases.
To test the framework under these completely random boundary conditions, the twin experiment is designed such that both the true and the baseline waves share the same target JONSWAP spectrum.
However, they are generated using distinct sets of random phases.
Consequently, although the two cases possess identical statistical characteristics, their instantaneous free surface elevations and velocity fields are uncorrelated.
The target spectrum has a peak period $T_p = 2.5$\,s and a significant wave height $H_s = 0.075$\,m. The signal is synthesized from 500 random wave components. The ensemble contains 50 members. Each member is also composed of 500 components with randomized phases. Observations are free surface elevations sampled at 20 locations that are uniformly distributed in the streamwise direction along the computational domain($L=22$~m). The observation noise standard deviation $\sigma_o$ is 0.002 m. The assimilation interval is 1 s. The modal truncation criterion remains set at the $99\%$ cumulative energy threshold.

In sequential assimilation, the filter remains effective only if the background error covariance retains a realistic magnitude. In practice, frequent updates and a finite ensemble can lead to ensemble collapse. The spread decreases too rapidly. The background error covariance then becomes underestimated. This causes weak corrections and can trigger filter divergence. Inflation is therefore required in sequential reconstruction. It restores ensemble diversity and sustains the corrective capability of the EnKF through long assimilation windows. 

\begin{figure}[htbp]
    \centering
    \includegraphics[width=0.7\textwidth]{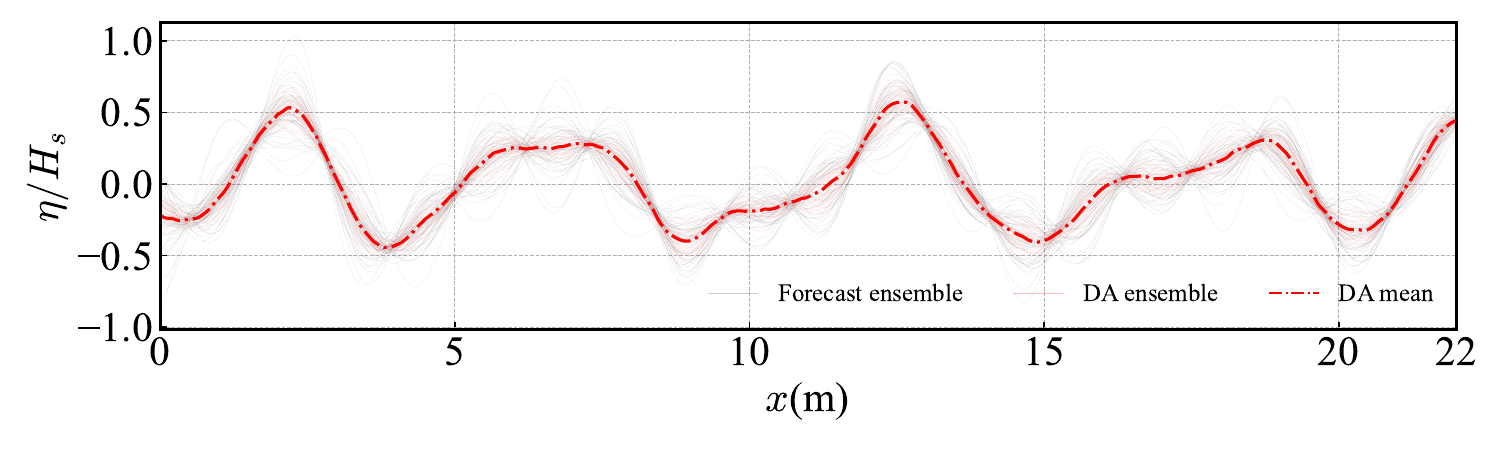}
    \caption{Wave ensemble profiles before and after inflation at the first assimilation time($t_{l_0}=8$~s).}
    \label{fig:inflation_surface}
\end{figure}

The effect of the surface inflation mentioned in Sec~\ref{sec:inflation} is illustrated by the free surface profiles before and after inflation in Fig~\ref{fig:inflation_surface}. The inflated ensemble is represented as the forecast ensemble(grey lines), and the DA ensemble is the elevations before inflation(red lines). The procedure increases ensemble spread in a controlled manner. It also preserves the spectral structure and avoids unphysical modifications of the gas-liquid topology. 

\begin{figure}[htbp]
\centering
\begin{subfigure}[b]{0.7\textwidth}
    \centering
    \includegraphics[width=\textwidth]{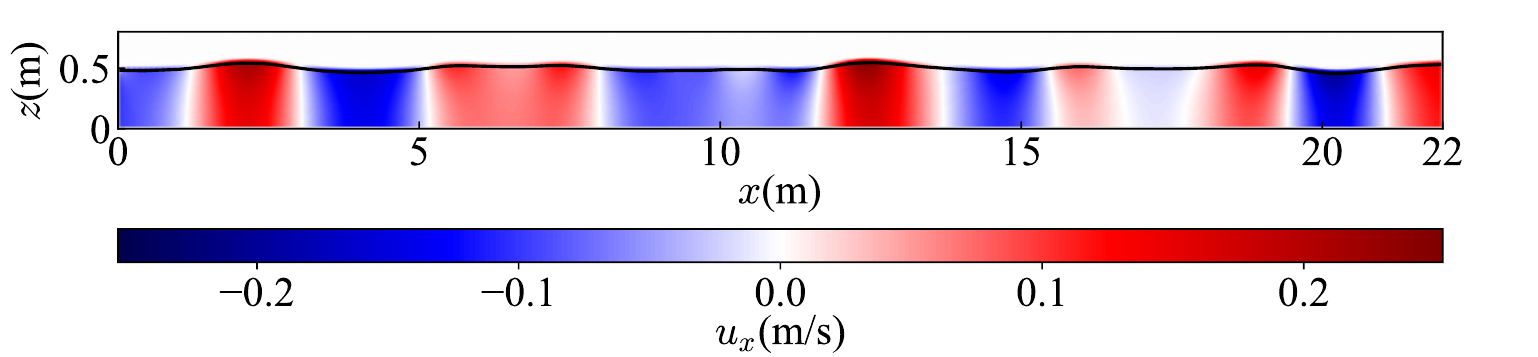}
    \caption{$u_x$ component}
\end{subfigure}\\

\begin{subfigure}[b]{0.7\textwidth}
    \centering
    \includegraphics[width=\textwidth]{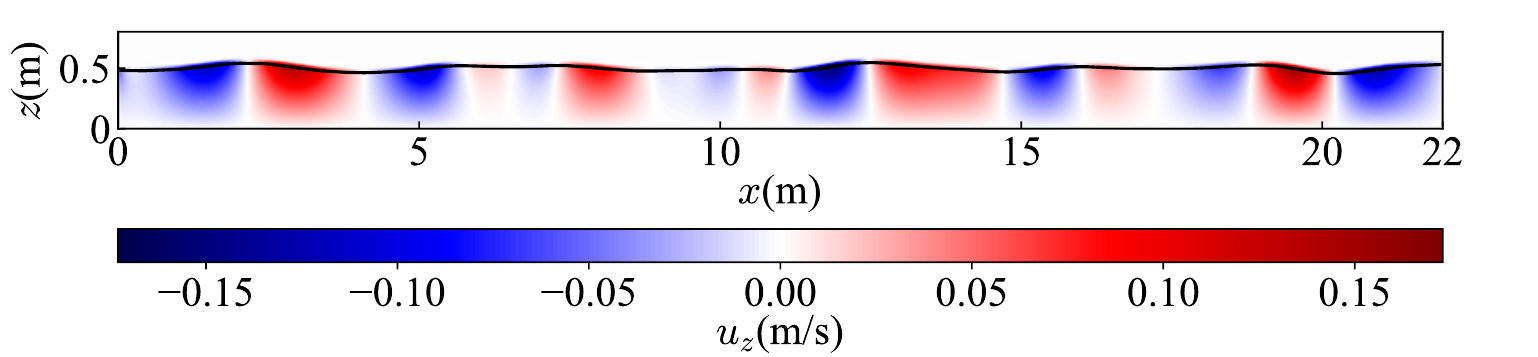}
    \caption{$u_z$ component}
\end{subfigure}\\

\caption{Inflated velocity components of a member based on the potential flow solution.}
\label{fig:inflation_velocity}
\end{figure}

Surface inflation alone is insufficient for the next forecast cycle. The velocity field is derived from a potential flow solution that is consistent with the inflated interface. It supplies a physically plausible kinematic field for the two-phase solver. It also reduces spurious transients at the restart of the forecast step. The resulting velocity field is shown in Fig~\ref{fig:inflation_velocity}.

The consistency is assessed by NEES. The best inflation coefficient is selected from this diagnostic. The NEES values and the corresponding waveform RMS are summarized in Table \ref{tab:nees_inflation}. Both ${\mathrm{NEES}}/r_{\phi}$ and $\epsilon_{RMS}^{\eta}$ are time-averaged over an evaluation window of four peak periods, i.e. $T=4T_p$. Table \ref{tab:nees_inflation} indicates that the inflation coefficient $\varrho=2.0$ provides the best overall performance. It yields $\overline{\mathrm{NEES}/r_{\phi}}=1.0990$, which is closest to unity among the tested values, suggesting the most consistent uncertainty quantification. It also achieves the minimum time-averaged waveform error, with $\overline{\epsilon_{RMS}^{\eta}}=6.13\times10^{-3}$ over $T=4T_p$. Therefore, $\varrho=2.0$ is adopted in the subsequent reconstructions.

\begin{table}[t]
\centering
\caption{Time-averaged NEES-based consistency results and waveform errors for different inflation coefficients. The averaging window is $T=4T_p$.}
\label{tab:nees_inflation}
\begin{tabular}{ccc}
\hline
$\varrho$ & $\overline{\mathrm{NEES}/r_{\phi}}$ & $\overline{\epsilon_{RMS}^{\eta}}$ \\
\hline
1.0 & 1.6052 & $7.19\times 10^{-3}$ \\
1.5 & 1.3692 & $6.95\times 10^{-3}$ \\
2.0 & 1.0990 & $6.13\times 10^{-3}$ \\
2.5 & 0.8360 & $7.27\times 10^{-3}$ \\
\hline
\end{tabular}
\end{table}

\begin{figure}[htbp]
\centering
\begin{subfigure}[b]{0.48\textwidth}
    \centering
    \includegraphics[width=\textwidth]{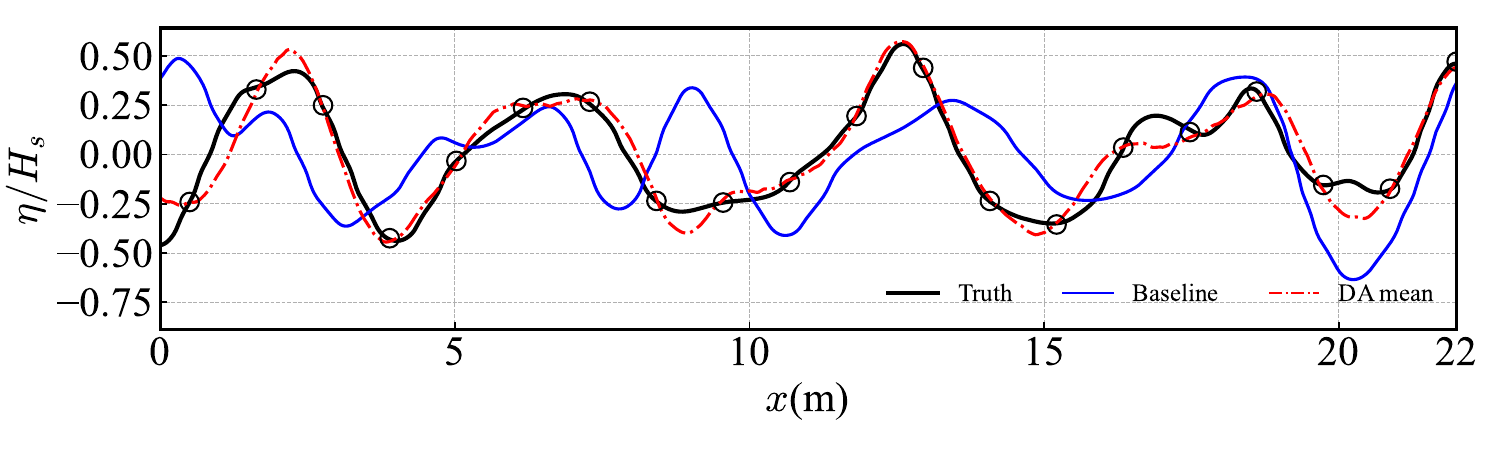}
    \caption{$t_{l_0}/T_p=0$}
\end{subfigure}
\hfill 
\begin{subfigure}[b]{0.48\textwidth}
    \centering
    \includegraphics[width=\textwidth]{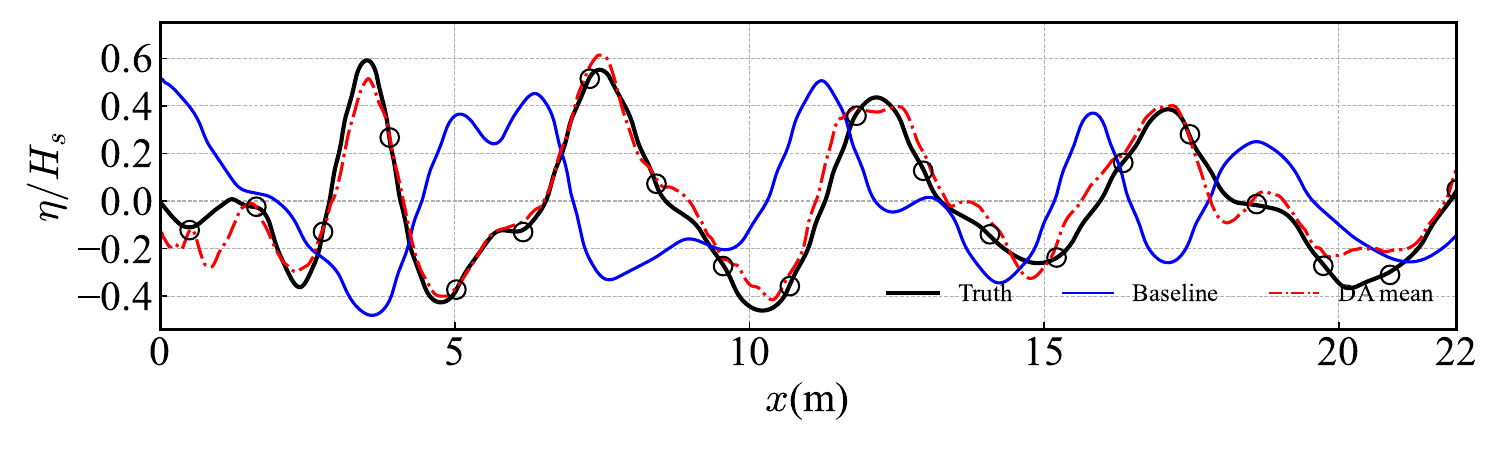}
    \caption{$t_{l_{5}}/T_p=2$}
\end{subfigure}

\vspace{0.2cm}

\begin{subfigure}[b]{0.48\textwidth}
    \centering
    \includegraphics[width=\textwidth]{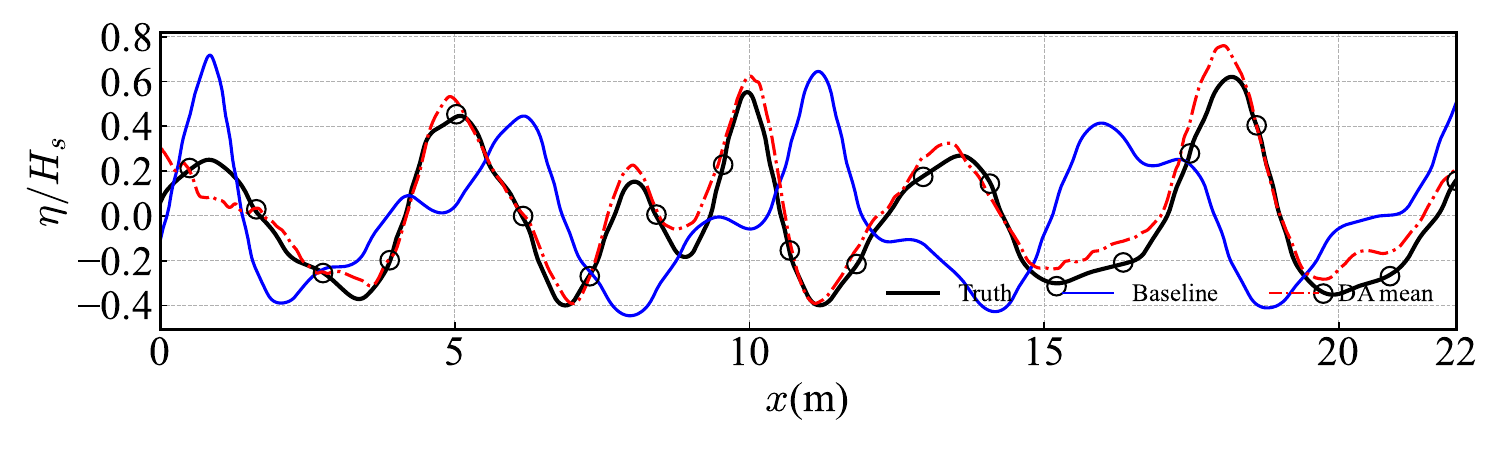}
    \caption{$t_{l_{10}}/T_p=4$}
\end{subfigure}
\hfill
\begin{subfigure}[b]{0.48\textwidth}
    \centering
    \includegraphics[width=\textwidth]{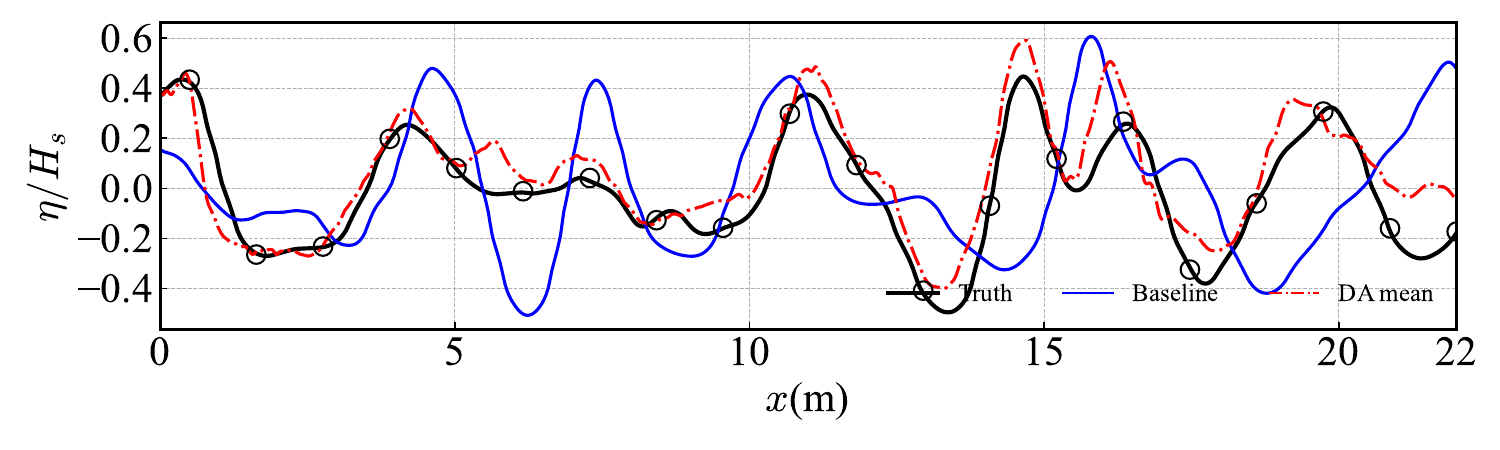}
    \caption{$t_{l_{15}}/T_p=6$}
\end{subfigure}

\vspace{0.2cm}

\begin{subfigure}[b]{0.48\textwidth}
    \centering
    \includegraphics[width=\textwidth]{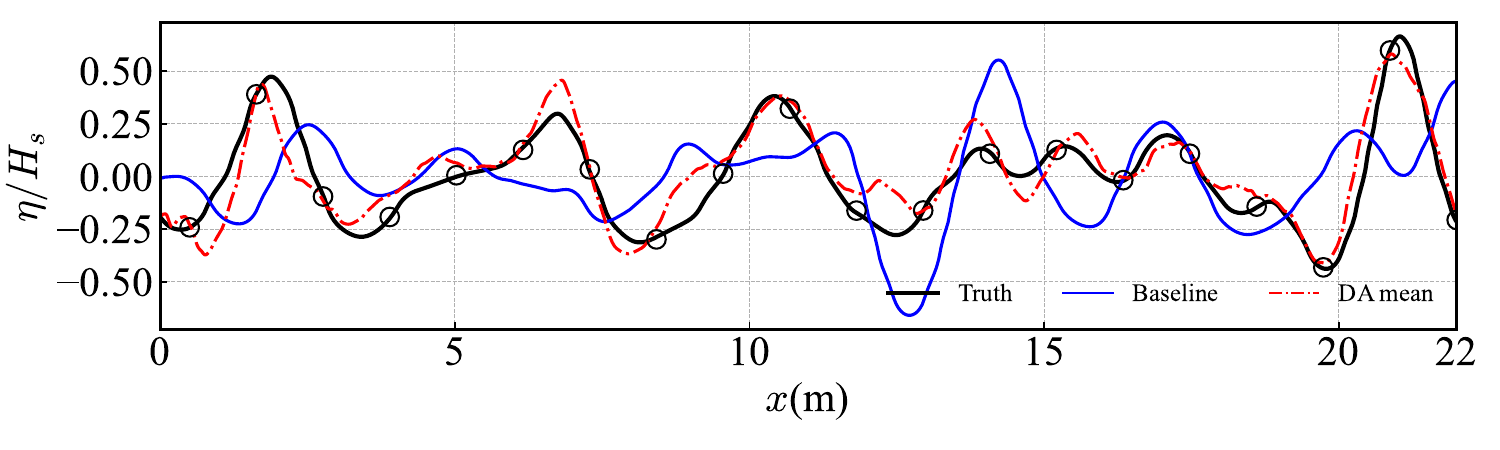}
    \caption{$t_{l_{20}}/T_p=8$}
\end{subfigure}
\hfill
\begin{subfigure}[b]{0.48\textwidth}
    \centering
    \includegraphics[width=\textwidth]{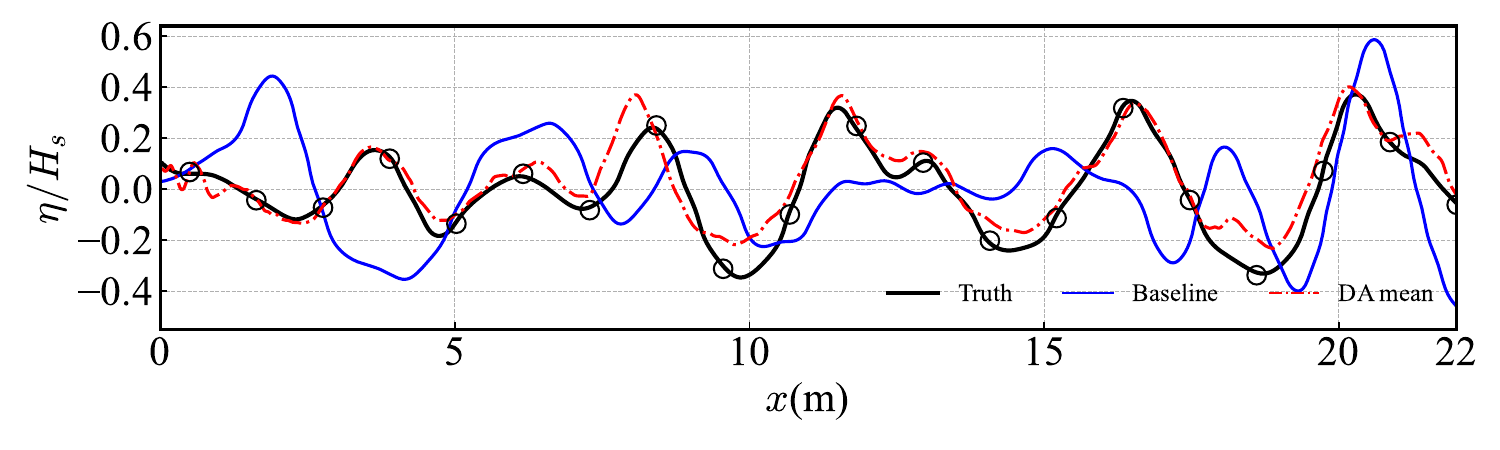}
    \caption{$t_{l_{25}}/T_p=10$}
\end{subfigure}

\caption{Surface elevations $\eta^a(x)$, $\eta^b(x)$, and $\eta^t(x)$ at different assimilation instants, cycles represent elevation observations.}
\label{fig:seq_surface}
\end{figure}

To assess the sequential assimilation performance, we compare the reconstructed free surface with the truth and the baseline at representative instants that cover different stages of the assimilation window. The curves in Fig~\ref{fig:seq_surface} include the DA mean, the truth, and the baseline. 

Fig~\ref{fig:seq_surface} shows that the DA mean $\eta^a(x)$ provides a substantial improvement over the baseline $\eta^b(x)$ at all instants. The reconstructed profiles are, in a global sense, close to the truth $\eta^t(x)$ across the full domain, which demonstrates that the framework supports long-time sequential assimilation for an irregular wave train. By contrast, the baseline remains inconsistent with the truth at all reported instants. This is expected because the inlet boundary condition is non-periodic and phase-resolved. The baseline and the truth share the same target spectrum but use different phase realizations. In this setting, the purpose of sequential assimilation is to continuously correct the phase-resolved wave field so that the reconstructed wave train matches the specific truth realization, rather than reproducing only the statistical wave properties. Despite the overall agreement, small local discrepancies can still be observed.
It is also observed that the reconstruction near the inlet can exhibit noticeable discrepancies. This is mainly caused by the ensemble wave generation, which introduces additional band components that differ from the truth realization. These components have not yet been sufficiently constrained by repeated assimilation cycles. As the wave train propagates downstream and successive analyses are performed, the sequential EnKF progressively filters out the spurious band content and aligns the phase-resolved wave field with $\eta^t(x, t)$. The downstream reconstruction therefore becomes increasingly optimized as more observations are assimilated.

\begin{figure}[htbp]
    \centering
    \includegraphics[width=0.7\linewidth]{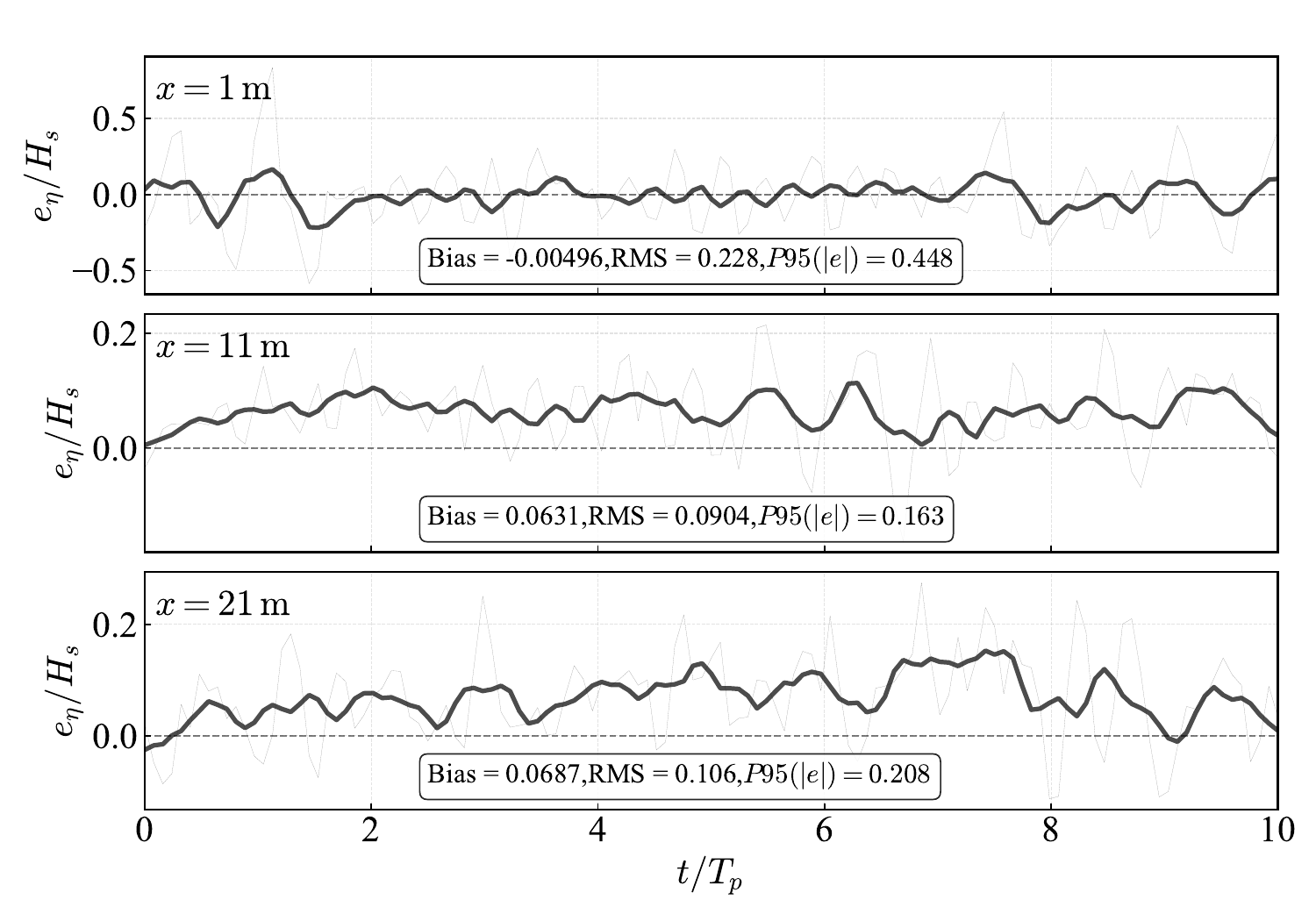}
    \caption{Time histories of nondimensional surface elevation errors at three gauges, with bias, RMS, and $P95(|e|)$.}
    \label{fig:sensor_error}
\end{figure}

The reconstruction error is quantified using the nondimensional elevation error normalized by $H_s$. The thin solid line in Fig~\ref{fig:sensor_error} reports the time history of this error at three gauge locations, together with the bias, RMS, and the $95\%$ quantile of the absolute error, denoted as $P95(|e|)$. The thick solid line shows a smoothed trend computed by a moving average. The averaging window spans half of the dominant wave period, i.e., $0.5T_p$. The trend highlights the global error level while filtering out short-time oscillations. At the downstream gauges $x=11\,\mathrm{m}$ and $x=21\,\mathrm{m}$, the error fluctuates around zero with a small positive bias of about $0.06$. The corresponding RMS values are $0.0904$ and $0.106$, and $P95(|e|)$ equals $0.163$ and $0.208$, respectively. In physical units, these RMS levels correspond to approximately $7$--$8\,\mathrm{mm}$, indicating that the reconstructed free surface is accurate in the main propagation region. 

\begin{figure}[h]
    \centering
    \includegraphics[width=0.7\linewidth]{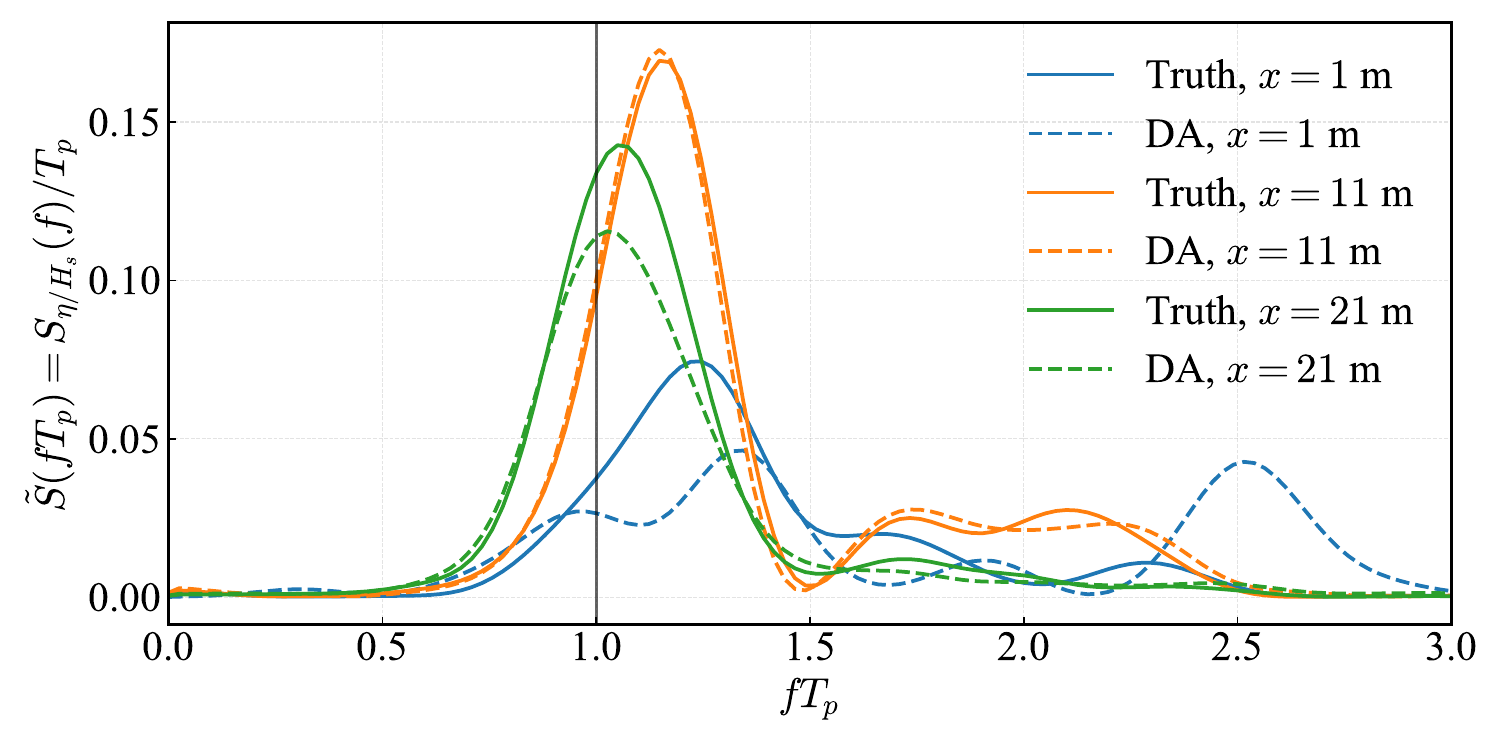}
    \caption{One-sided nondimensional power spectral density of surface elevation at gauges.}
    \label{fig:gauge_spectra}
\end{figure}

Fig~\ref{fig:gauge_spectra} presents the one-sided power spectral density (PSD) estimates of the nondimensional surface elevation time series $\eta/H_s$ at three gauges. The spectra are nondimensionalized as $\tilde S(fT_p)=S_{\eta/H_s}(f)/T_p$, such that the integral over the nondimensional frequency $\xi=fT_p$ satisfies $\int_{0}^{\infty}\tilde S(\xi)\,\mathrm{d}\xi=\mathrm{Var}(\eta/H_s)$. The vertical line at $\xi=1$ corresponds to the peak frequency $f_p=1/T_p$, which represents the dominant energy-containing scale of the irregular wave.

At the gauges located in the main propagation region ($x=11~\mathrm{m}$ and $x=21~\mathrm{m}$), the reconstructed spectra (DA) closely match the truth in the primary energy band. Both spectra exhibit a dominant peak centered around $\xi\approx 1$, with only minor differences in peak magnitude and bandwidth. This agreement indicates that the reconstruction preserves the energy distribution near the spectral peak. In addition, a secondary energy bulge appears on the high-frequency side in the range $\xi\approx 1.6$--$2.3$, which may be attributed to higher-frequency components of the irregular wave train. The DA spectrum reproduces this secondary feature well at $x=11~\mathrm{m}$, while at $x=21~\mathrm{m}$ the overall trend remains consistent but the magnitude is slightly lower. 

In contrast, the gauge at $x=1~\mathrm{m}$ shows noticeably reduced spectral agreement. 
Differences are evident around the primary peak, and an additional pronounced energy peak appears near $\xi\approx 2.5$ in the DA spectrum, while the true spectrum remains comparatively low in that band. Such an artificial amplification of high-frequency energy would manifest in the time domain as enhanced short-scale oscillations, consistent with the larger RMS errors previously observed at this gauge. 
Given the proximity of this location to the inlet region, the spurious high-frequency content in the DA spectrum is more plausibly attributed to a boundary-consistency issue. In the present setup, the inlet boundary condition is not included in the state vector and is therefore not corrected consistently with the internal field. As a result, the analysed wave field can be locally incompatible with the prescribed inflow wave. This mismatch introduces a non-smooth transition, which effectively acts as a high-frequency disturbance source in the DA spectrum. 

Overall, using spectral consistency as a reconstruction-quality criterion, the present framework performs robustly in the main propagation region, while further improvements are warranted near the inlet.

\subsection{Plunging Wave Reconstruction}
\label{sec:plunging}

The plunging wave case is selected to test the reconstruction framework in a regime where the downstream dynamics are strongly nonlinear. Wave shoaling on a mild slope leads to rapid steepening and eventual overturning. The resulting plunging wave is highly sensitive to small errors in amplitude and phase. It also involves strong viscous dissipation and complex two-phase interface motion. These characteristics make plunging waves a stringent benchmark for any reconstruction method, because modest mismatches before running up can shift the breaking onset time and the breaker location.

It is important to note that data assimilation is not performed during the breaking process. The assimilation window is restricted to the upstream constant depth region, where the interface topology is well-behaved. This choice avoids direct updates of an overturning interface and corresponding velocity. The objective is instead to reconstruct a incident wave before shoaling. The reconstructed wave then propagates toward the slope under the VOF-based CFD model without further correction. This design improves the reliability of the predicted wave features and still demonstrates that the framework can handle strongly nonlinear conditions.

\begin{figure}[htbp]
\centering
\includegraphics[width=0.9\textwidth]{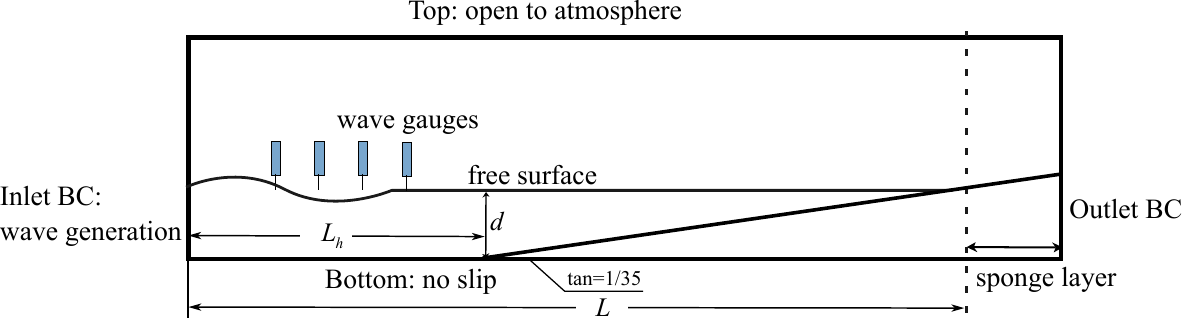}
\caption{Plunging wave numerical tank setup.}
\label{fig:plunging setup}
\end{figure}

The numerical setup follows a classical plunging wave experiment configuration, as shown in Fig~\ref{fig:plunging setup}. The toe of the slope is preceded by a constant depth region of length $L_h = 10~\mathrm{m}$. The still-water depth is $d = 0.4~\mathrm{m}$. The truth and baseline waves are cosine waves with the same period $T = 5~\mathrm{s}$. The truth wave height is $H^{t} = 0.12~\mathrm{m}$, while the baseline wave height is $H^{b} = 0.09~\mathrm{m}$. The ensemble size is $N = 50$. The ensemble is generated by perturbing the wave height with Gaussian noise with standard deviation $\epsilon = 0.5H^{b}$. Free surface elevation observations are collected at 4 wave gauges uniformly distributed on the constant depth region with spacing $\Delta x = 1$~s. The observation noise standard deviation is $\sigma_o = 0.002~\mathrm{m}$. The modal truncation criterion remains set at the $99\%$ cumulative energy threshold.

\begin{figure}[h]
\centering
\begin{subfigure}[b]{0.5\textwidth}
    \centering
    \includegraphics[width=\textwidth]{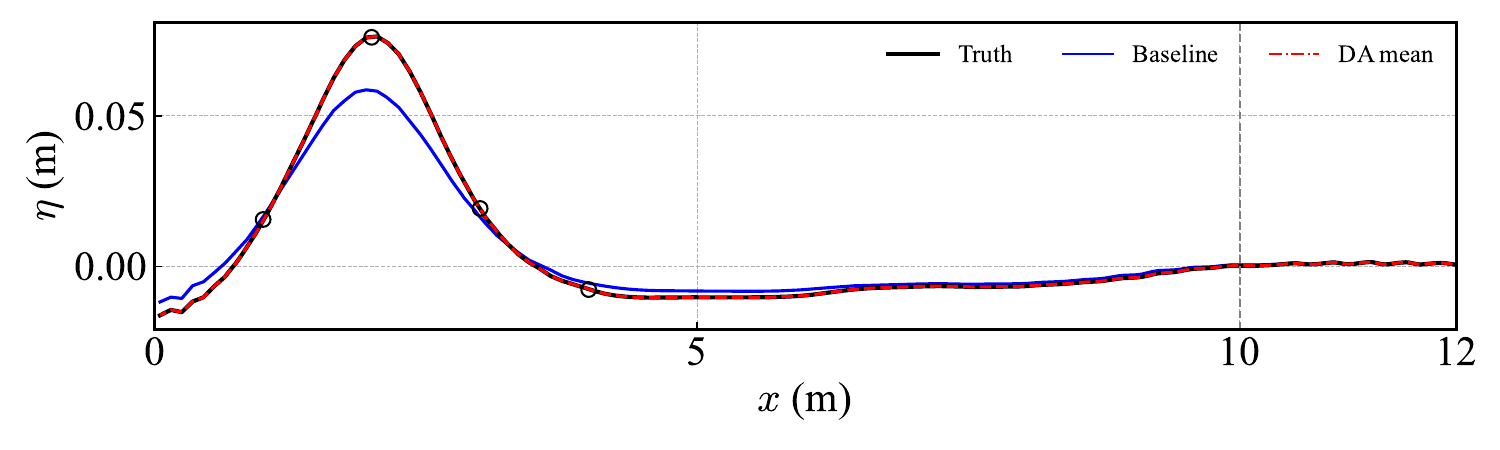}
    \caption{$t_{l_0}=6s$}
    \label{fig:height_6s}
\end{subfigure}\\

\begin{subfigure}[b]{0.5\textwidth}
   \centering
    \includegraphics[width=\textwidth]{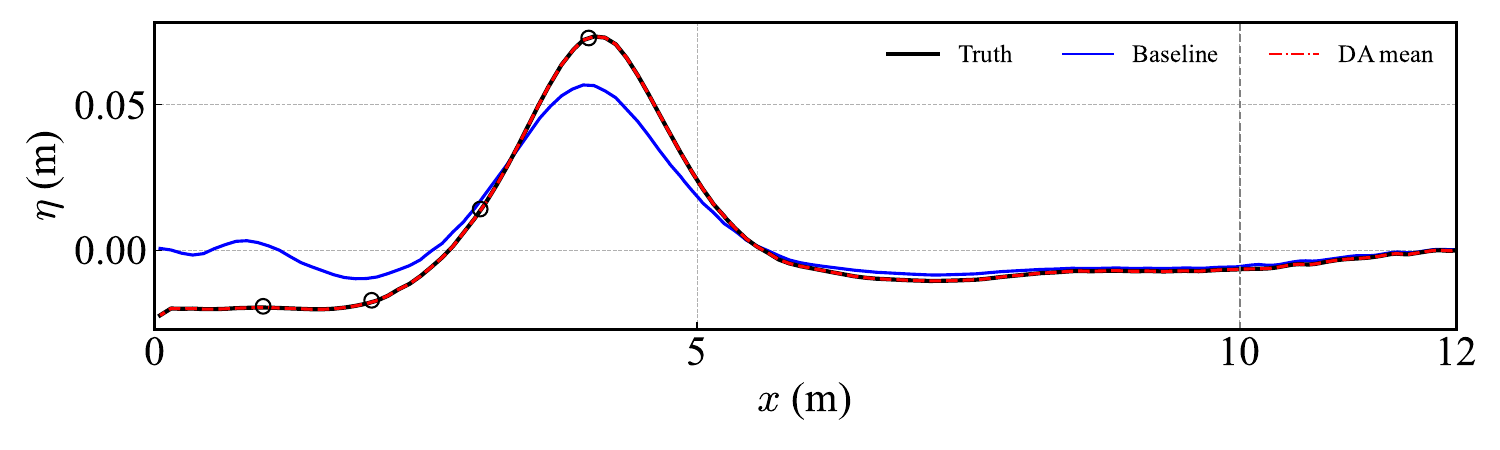}
    \caption{$t_{l_1}=7s$}
    \label{fig:height_7s}
\end{subfigure}\\

\begin{subfigure}[b]{0.5\textwidth}
   \centering
    \includegraphics[width=\textwidth]{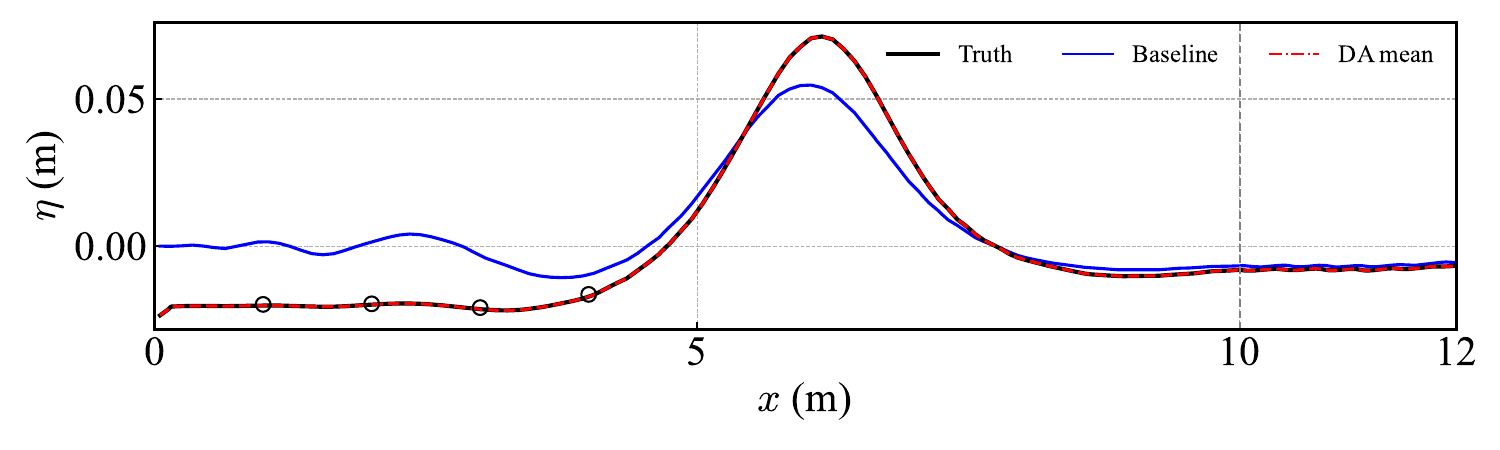}
    \caption{$t_{l_2}=8s$}
    \label{fig:height_8s}
\end{subfigure}\\

\caption{Surface elevations $\eta^b(x)$, $\eta^t(x)$ and $\eta^a(x)$ at the assimilation instants, cycles represent elevation observations.}
\label{fig:plunging DA elevations}
\end{figure}

The performance of the proposed framework is evaluated by examining the free surface evolution at three consecutive assimilation instants: $t_{l_0}=6$~s, $t_{l_1}=7$~s, and $t_{l_2}=8$~s.
Fig~\ref{fig:plunging DA elevations} compares the wave profiles of the true state, the baseline simulation, and the DA analysis mean.
The computational domain shown covers the constant depth region ($x \le 10$~m) and the beginning of the slope.

It is important to note that despite identical settings for the wave period and the start time, a slight phase discrepancy exists between the baseline and the true state.
This misalignment originates from the wave generation boundary condition, where a ramp time of $6$~s is applied.
While both simulations generate a fully developed wave within this duration, the specific phase evolution during this transient ramp-up period is not guaranteed to be identical. 
At $t_{l_0}=6$~s (Fig~\ref{fig:height_6s}), consistent with the experimental setup, the baseline simulation (blue line) significantly underpredicts the wave amplitude.
Nevertheless, the DA update (red dash-dotted line) effectively corrects both deficiencies.
By assimilating the sparse gauge data, the reconstructed wave height is adjusted to match the target $H^t = 0.12$~m, and the phase is perfectly aligned with the true state.
This high-fidelity reconstruction is maintained as the wave propagates forward.
At $t_{l_1}=7$~s (Fig~\ref{fig:height_7s}) and $t_{l_2}=8$~s (Fig~\ref{fig:height_8s}), the wave train advances towards the slope.
Throughout this process, the DA analysis mean consistently tracks the ground truth.

\begin{figure}[]
\centering

\begin{subfigure}[b]{0.47\textwidth}
    \centering
    \includegraphics[width=\textwidth]{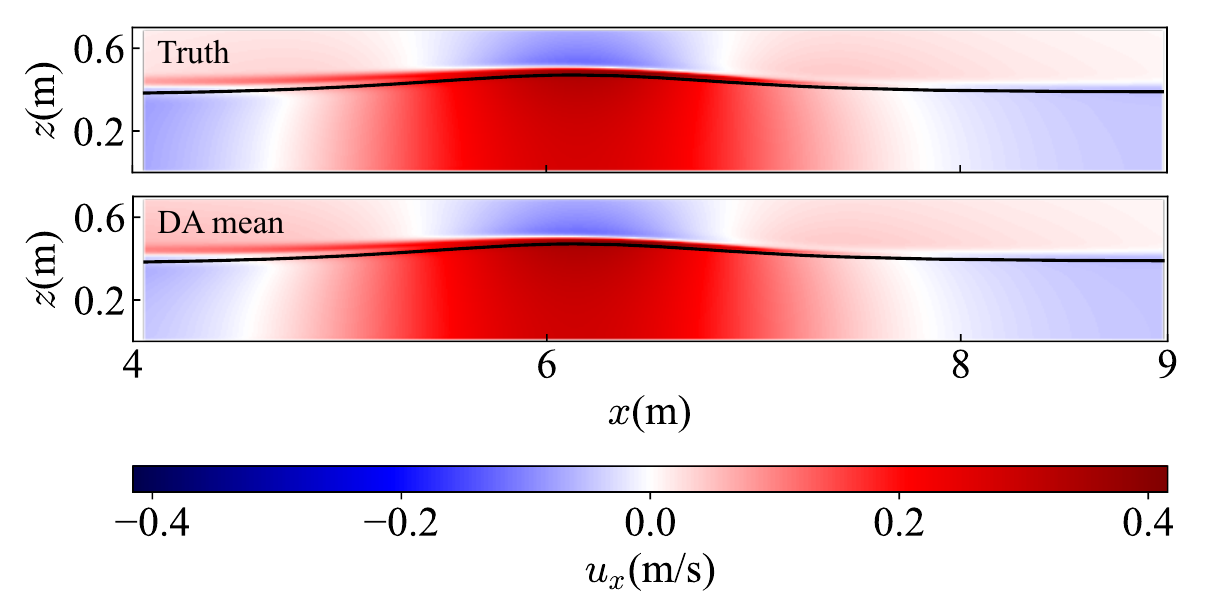}
    \caption{${u}_x$ comparison}
    \label{fig:plunging ux}
\end{subfigure}
\hfill
\begin{subfigure}[b]{0.47\textwidth}
    \centering
    \includegraphics[width=\textwidth]{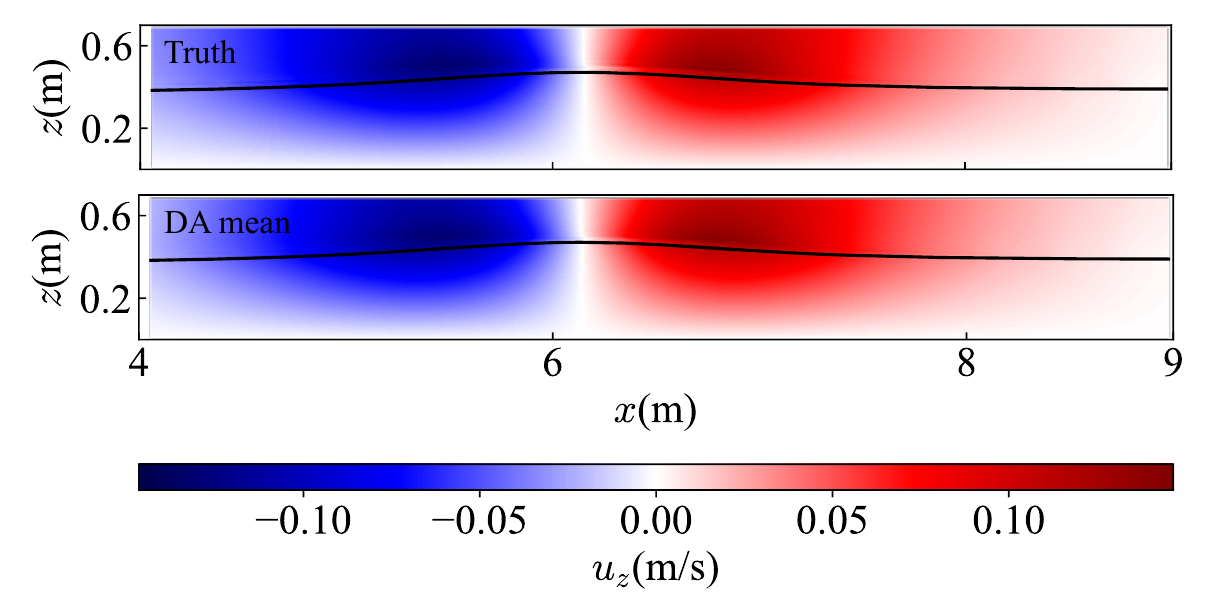}
    \caption{${u}_z$ comparison}
    \label{fig:plunging uz}
\end{subfigure}

\caption{Comparison of reconstructed and true velocity fields at the last assimilation instant $t_{l_2}=8$~s.}
\label{fig:plunging velocity}
\end{figure}

Beyond the free surface elevation, the accurate retrieval of the subsurface velocity field is crucial for the correct dynamic evolution of the subsequent shoaling process.
Fig~\ref{fig:plunging ux} and \ref{fig:plunging uz} present the comparison of the horizontal velocity $\boldsymbol{u}_x$ and the vertical velocity $\boldsymbol{u}_z$ between the true state and the DA analysis mean at $t=8$~s.
The plotted region corresponds to the constant depth zone ($x \in [4, 9]$~m) just before the slope.

The true state exhibits the classical motion with positive velocities concentrated under the wave crest (around $x \approx 6$~m) and negative velocities under the trough.
The DA reconstructed field captures this spatial distribution with remarkable accuracy.
Both the magnitude and the vertical gradient of $\boldsymbol{u}_x$ in the analysis mean are visually indistinguishable from the truth.
This indicates that the ensemble covariance successfully correlates the surface elevation innovation with the underlying horizontal velocity.

Similar agreement is observed in the vertical velocity $\boldsymbol{u}_z$, shown in Fig~\ref{fig:plunging uz}.
The vertical velocity field is characterized by a phase shift relative to the wave elevation: upward flow is observed at the wave front and downward flow at the wave rear.
The assimilation framework correctly reproduces this phase relationship.

The successful reconstruction of the velocity field at $t=8$~s confirms that the updated state vector contains a physically consistent set of flow variables. 
By accurately reconstructing the incident wave condition in the linear regime, the solver is able to correctly propagate the wave downstream into the shoaling region.
This ensures that when the simulation continues beyond the assimilation window into the shoaling region, the wave possesses not only the correct shape but also the correct kinetic energy distribution to undergo realistic steepening and breaking.

\begin{figure}[h]
\centering
\begin{subfigure}[b]{0.48\textwidth}
\centering
\includegraphics[width=\textwidth]{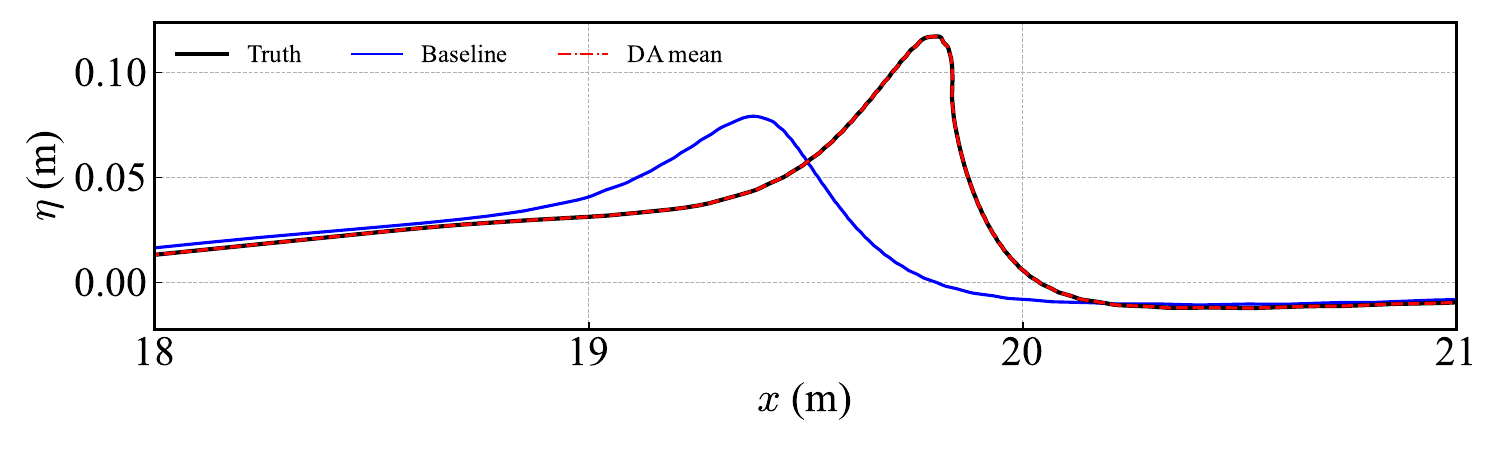}
\caption{$t=15.1~\mathrm{s}$}
\label{fig:plunging_forward a}
\end{subfigure}
\hfill
\begin{subfigure}[b]{0.48\textwidth}
\centering
\includegraphics[width=\textwidth]{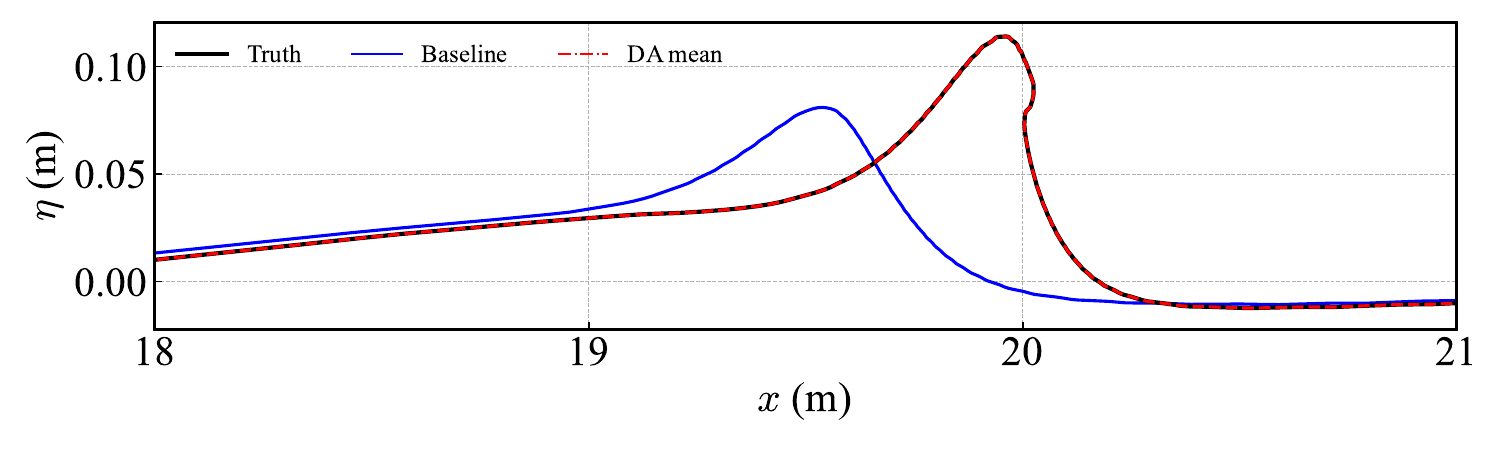}
\caption{$t=15.2~\mathrm{s}$}
\label{fig:plunging_forward b}
\end{subfigure}

\vspace{0.15cm}

\begin{subfigure}[b]{0.48\textwidth}
\centering
\includegraphics[width=\textwidth]{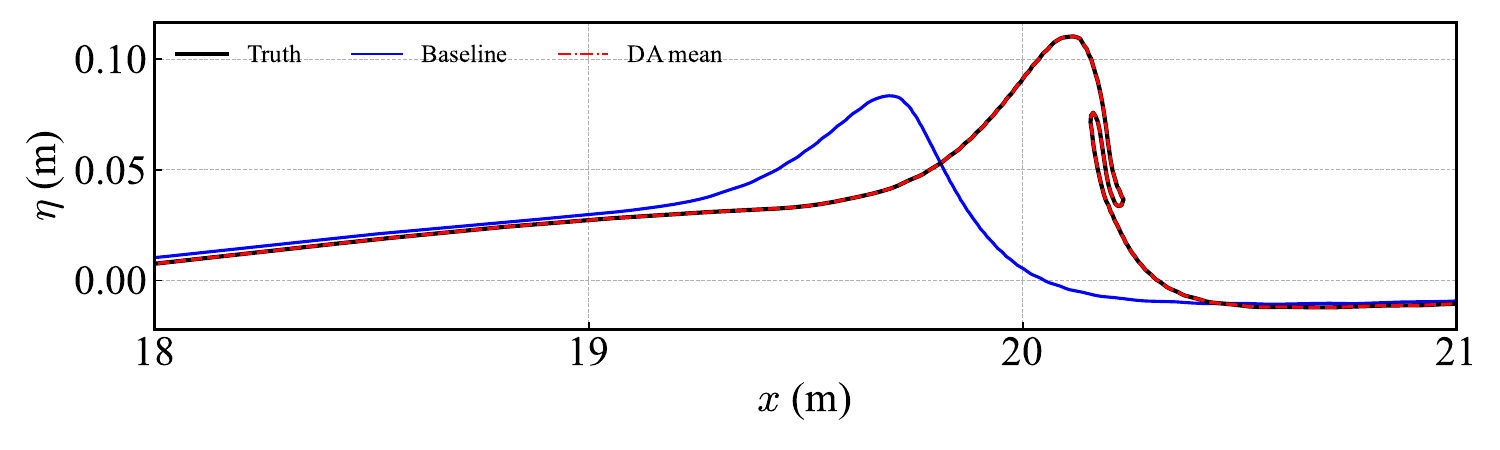}
\caption{$t=15.3~\mathrm{s}$}
\label{fig:plunging_forward c}
\end{subfigure}
\hfill
\begin{subfigure}[b]{0.48\textwidth}
\centering
\includegraphics[width=\textwidth]{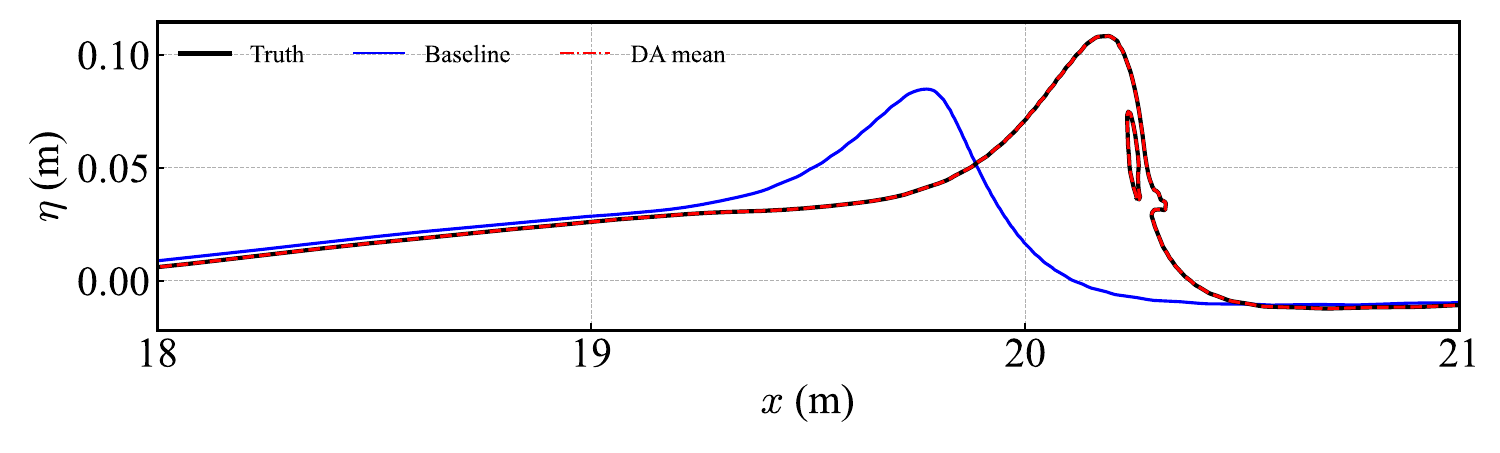}
\caption{$t=15.35~\mathrm{s}$}
\label{fig:plunging_forward d}
\end{subfigure}
\caption{Free surface elevations $\eta^a(x)$, $\eta^b(x)$, and $\eta^t(x)$ during the downstream shoaling and plunging stages.}
\label{fig:plunging_forward}
\end{figure}

A key advantage of the proposed framework is its ability to predict downstream nonlinear behaviors by correctly initializing the upstream wave.
To validate this, we examine the wave evolution on the slope long after the assimilation has ceased.
Fig~\ref{fig:plunging_forward} displays the frame-by-frame comparison of the free surface elevation during the breaking process, from $t=15.1$~s to $t=15.35$~s.
The region shown corresponds to the surf zone ($x \in [18, 21]$~m), where the water depth is shallow and the nonlinearity is strongest.

The sequence captures the rapid topological changes characteristic of a plunging breaker.
At $t=15.1$~s (Fig~\ref{fig:plunging_forward a}, the wave front becomes nearly vertical due to the shoaling effect.
The baseline simulation (blue line), suffering from the initial amplitude and phase errors, fails to reach the critical steepness required for breaking at this location.
In contrast, the DA analysis mean (red dash-dotted line) accurately reproduces the steep wave front of the true state (black line).

As the wave propagates to $t=15.2$~s (Fig~\ref{fig:plunging_forward b}), the crest begins to overturn.
The DA result captures this overturning initiation overlapping well with the truth.
The baseline, however, lags significantly behind and exhibits a different crest shape.

Remarkably, even in this highly nonlinear stage involving complex interface deformation, the reconstructed wave remains consistent with the ground truth as shown in Fig~\ref{fig:plunging_forward c} and \ref{fig:plunging_forward d}.
This result confirms that by assimilating data in the linear regime upstream, the EnKF successfully infers the correct flow field, enabling the VOF solver to naturally and accurately evolve the wave into a breaking state without further corrections.

\section{Conclusion}
\label{sec:conclusion}

In this study, we have developed and validated a sequential data assimilation framework for wave reconstruction in VOF-based numerical wave tanks.
By integrating the EnKF with a finite volume Navier-Stokes solver, the proposed framework addresses the challenge of reconstructing specific deterministic wave realizations from limited surface elevation measurements.
POD was employed to reduce the dimensionality of the high-fidelity state space, making the ensemble-based assimilation computationally feasible.

Investigation into interface representation revealed that converting the discontinuous VOF fraction into a level-set field significantly improves the compactness of the modal basis.
Furthermore, a physics-constrained inflation strategy was introduced.
By inflating the spectral amplitudes of the free surface and updating the velocity field via potential flow theory, the framework successfully mitigated ensemble collapse and maintained filter consistency during long-term sequential updates.

The framework was tested against three distinct wave conditions, including regular waves, irregular waves, and plunging waves. 
A distinguishing feature of the proposed framework, compared to traditional potential flow-based methods, is its capability to capture strongly nonlinear flow phenomena through high-fidelity CFD modeling.
This was demonstrated in the plunging wave case, where the assimilation of linear waves in the upstream region allowed for the accurate prediction of the downstream overturning and breaking process.
The solver naturally evolved the reconstructed states into complex breaking topologies.

In addition to handling nonlinear events, the framework exhibited robust performance in the phase-resolved reconstruction of irregular waves.
It successfully adapted to the stochastic nature of random wave trains by sequentially assimilating measurement data.
The method effectively corrected the phase and amplitude of the instantaneous wave field to match the specific realization of the truth.
This confirms the framework's ability to adapt to continuously changing wave conditions and eliminate errors arising from boundary mismatches or initial uncertainties.

In summary, this work establishes a feasible pathway for constructing ``digital twin'' wave tanks capable of synchronizing with physical reality.
Future efforts will focus on three key directions to further advance this framework.
First, we aim to accelerate the computational efficiency of the framework to achieve real-time assimilation, which is a prerequisite for live digital twinning.
Second, the methodology will be validated against physical wave tank experiments to bridge the gap between numerical simulations and laboratory environments.
Finally, the framework will be extended to reconstruct three-dimensional wave fields, enabling the analysis of more complex wave-structure interactions in ocean engineering applications.

\section*{Acknowledgments}
This work is supported by the Research Project of China COSCO Shipping Corporation Limited (2023-2-Z001-03), Open Project of the State Key Laboratory of Ocean Engineering, and the Fundamental Research Funds for the Central Universities. 

\appendix
\section{Observation Operator}
\label{sec:obervation operator}

In this study, different representations of the free surface are employed, associated with different observation operators. 

First, we derive the observation operators corresponding to the different free surface representations when adopting free surface elevations as observations. 
The mode eigenvalue of the phase field for ensemble member $n$ is represented as $a^{(n), f}$. The modal-to-physical reconstructions are
\begin{equation}
    \boldsymbol{\phi}^{(n),f} = \hat{\boldsymbol{U}}_{\phi}\,\boldsymbol{a}^{(n),f}.
\end{equation}

Note the observation operator acting on the state can be written as
\begin{equation}
    \mathcal{H}(\boldsymbol{a}^{(n), f}) = 
\boldsymbol{S}\,\mathcal{F}\!\left(\boldsymbol{\phi}^{(n), f} \right),
\end{equation}
where $\boldsymbol{S}$  belongs to $\mathbb{R}^{p_k\times g}$ is the interpolation operator that extracts elevation values at the observation locations;
$\mathcal{F}:\mathbb{R}^{M}\to\mathbb{R}^g$ is an operator that converts the implicit phase field to a set of nodes representing the geometric free surface elevation in the entire region, where $g$ is the number of nodes.

In the mathematical formulation presented above, the level-set field $\boldsymbol{\phi}$ is used as the representative state variable for the free surface.
It is important to note that the same derivation and logic apply analogously to the VOF field $\boldsymbol{\alpha}$.

\bibliographystyle{elsarticle-num-names} 
\bibliography{ref}







\end{document}